\documentclass[a4paper]{article}
\usepackage{a4wide}
\usepackage{amssymb}
\usepackage[english]{babel}
\usepackage[dvips]{graphicx}
\usepackage{xspace}
\newcommand{\ket}[1]{\ensuremath{|\,#1\,\rangle}}

\newcommand{\qram}{{\em QRAM}\xspace}
\newcommand{\lang}[1]{\texttt{#1}\xspace}
\newcommand{\cpp}{\lang{C++}}
\newcommand{\pgcl}{\lang{pGCL}}
\newcommand{\qgcl}{\lang{qGCL}}
\newcommand{\qcl}{\lang{QCL}}
\newcounter{Lcount}
\newenvironment{myenum}{\begin{list}{\fbox{\arabic{Lcount}}}
    {\usecounter{Lcount}}}
  {\end{list}}
\newcommand{\myitem}[1]{\item \underline{\em #1} \par}
\newcommand{\myex}[2]{
  $\blacktriangleright$
  \parbox{.48\linewidth}{\mbox{\texttt{#1}}}
  \parbox{.48\linewidth}{\mbox{#2}}}
\newcommand{\mydagger}{\raisebox{2pt}[0pt][0pt]{\makebox[0pt]{$\dagger$}}}
\newsavebox{\picinbox}
\newlength{\picinboxlen}
\newlength{\picinboxrest}
\newcommand{\picinpar}[2]{%
  \sbox{\picinbox}{#2}%
  \settowidth{\picinboxlen}{\usebox{\picinbox}}%
  \setlength{\picinboxrest}{\linewidth}%
  \addtolength{\picinboxrest}{-\picinboxlen}%
  \addtolength{\picinboxrest}{-3pt}%
  \begin{minipage}[t]{\picinboxrest} #1 \end{minipage}
  \begin{minipage}[t]{\picinboxlen}
    \rule{\baselineskip}{0pt} \vspace{-\baselineskip} \\
    \usebox{\picinbox} 
  \end{minipage} \par
}
\usepackage{texdraw}    
\input{txdtools}        
\usepackage{ifthen}     

\newcommand{\qdrawdim}{pt}
\newcommand{\qlinewidth}{.5}
\newcommand{\qcircuitcalc}[3]{\drawdim{\qdrawdim} \linewd{\qlinewidth}
                              \realmult{#1}{1}          \step
                              \realadd {#2}{1}          \slots
                              \realmult{\step}{.5}      \halfstep
                              \realmult{\step}{.7}      \ministep
                              \realmult{\ministep}{.5}  \halfministep
                              \realmult{\ministep}{.3}  \ballradius
                              \realmult{\step}{.13}     \dotradius
                              \realmult{\dotradius}{1.} \multilineskip
                              \realmult{\step}{\slots}  \tot
                              \realmult{#3}{1}          \text
                              \realadd {\tot}{\text}    \ttot
                              \realadd {0}{1}           \theslot
                              \realmult{.9}{1}          \boxfill
                             }
\newcounter{line_counter}
\newenvironment{qcircuit}[4]{\begin{tabular}[c]{c}\begin{texdraw}
                             \qcircuitcalc{#1}{#2}{#4}
                             \setcounter{line_counter}{#3}
                             \whiledo{\value{line_counter} > 0}
                             {\addtocounter{line_counter}{-1}
                              \qline{\value{line_counter}}
                             }
                            }
                            {\end{texdraw}\end{tabular}}

\newcommand{\qsame}[1][1]{\realadd{\theslot}{-#1}\theslot}
\newcommand{\qskip}[1][1]{\realadd{\theslot}{#1}\theslot}
\newcommand{\qcalcpos}[1]{\realmult{#1}{\step}          \vert
                          \realmult{\step}{4}           \vertoffset
                          \realadd{\vert}{\vertoffset}  \vert 
                          \realmult{\theslot}{\step}    \hori
                          \realadd {\hori}{\text}       \thori}
\newcommand{\qboxspace}{\rmove({-\halfministep} {-\halfministep})
                        \rmove({\ministep} 0)  \rmove(0 {\ministep})
                        \rmove({-\ministep} 0) \rmove(0 {-\ministep}) 
                        \rmove({\halfministep} {\halfministep})}
\newlength{\qtextlength}
\newcommand{\qtextL}[1]{\textref h:L v:C \htext{#1} \drawdim{sp}
                        \settowidth{\qtextlength}{#1}
                        \rmove({\number\qtextlength} 0) \drawdim{\qdrawdim}}
\newcommand{\qtextR}[1]{\textref h:R v:C \htext{#1} \drawdim{sp}
                        \settowidth{\qtextlength}{#1}
                        \rmove({-\number\qtextlength} 0) \drawdim{\qdrawdim}}
\newcounter{domultilines}
\newcommand{\qgenerateline}[6]{\qcalcpos{#1} \move({\text} {\vert})
                               \realmult{#2}{\step}\frompatch
                               \realmult{#3}{\step}\topatch
                               \realadd{\topatch}{-\frompatch}\topatch
                               \setcounter{domultilines}{#5}
                               \realadd{#5}{-1}\vshift
                               \realmult{\multilineskip}{\vshift}\vshift
                               \realmult{\vshift}{0.5}\vshift
                               \rmove({\frompatch} {\vshift}) 
                               \whiledo{\value{domultilines} > 0}
                                    {\addtocounter{domultilines}{-1}
                                     \ifthenelse{\equal{#6}{del}}
                                        {\linewd 1 \setgray 1
                                         \rlvec({\topatch} 0)
                                         \linewd {\qlinewidth} \setgray 0
                                         \rmove({-\topatch} 0)}{}
                                     \lpatt(#4) \rlvec({\topatch} 0)
                                     \lpatt() \rmove({-\topatch} 0)
                                     \rmove(0 {-\multilineskip})
                                    }
                               \rmove(0 {\vshift}) \rmove(0 {\multilineskip})
                               }
\newcommand{\qmultiline}[2]{\qgenerateline{#1}{0}{\slots}{}{#2}{del}}
\newcommand{\qline}[1]{\qgenerateline{#1}{0}{\slots}{}{1}{} \qboxspace}
\newcommand{\qpatchline}[4]{\qgenerateline{#1}{#2}{#3}{#4}{1}{del}}

\newcommand{\qlabel}[2]{\qcalcpos{#1} \move(0 {\vert})
                        \textref h:L v:C \htext{#2}}

\newcommand{\qendlabel}[2]{\qcalcpos{#1} \move({\ttot} {\vert})
                           \rmove({\halfministep} 0) \qtextL{#2}}

\newcommand{\qgenericbox}[5]{\qcalcpos{#1} \move({\thori} {\vert})
                             \realadd{#2}{-#1}\thelinerange
                             \realmult{\thelinerange}{\step}\maxistep
                             \realmult{#3}{1}\boxstep
                             \realmult{\boxstep}{.5}\halfboxstep
                             \realadd{\maxistep}{\boxstep}\maxistep
                             \rmove({-\halfboxstep} {-\halfboxstep})
                             \setgray #4
                             \rlvec({\boxstep} 0) \rlvec(0 {\maxistep})
                             \rlvec({-\boxstep} 0) \rlvec(0 {-\maxistep}) 
                             \lfill f:{\boxfill} \setgray 0
                             \ifthenelse{#5 < 0}{}
                                        {\textref h:L v:B \htext{\tiny {#5}}}
                             \realmult{\maxistep}{.5}\halfmaxistep
                             \rmove({\halfboxstep} {\halfmaxistep})
                            }
\newcommand{\qbox}[2]{\qgenericbox{#1}{#2}{\ministep}{0}{-1}}
\newcommand{\qevidence}[3][{}]{\setcounter{line_counter}{#3}
                               \whiledo{\value{line_counter} > 0}
                                       {\addtocounter{line_counter}{-1}
                                        \ifthenelse{\equal{#1}{}}
                                {\qgenericbox{#2}{#2}{\step}{0}{-1}}
                                {\qgenericbox{#2}{#2}{\step}{.8}{-1}}
                                        \qskip
                                       }
                               \qskip[-#3] 
                               \realmult{\step}{#3} \textoffset
                               \realadd{\textoffset}{-\step} \textoffset
                               \realmult{\textoffset}{.5} \textoffset
                               \rmove({-\textoffset} 0)
                               \textref h:C v:C \htext{#1}
                              }

\newcommand{\qbaresegment}[2]{\qcalcpos{#1} \move({\thori} {\vert})
                              \qcalcpos{#2} \lvec({\thori} {\vert})}
\newcommand{\qsegment}[2]{\qcalcpos{#2} \move({\thori} {\vert})
                          \fcir f:0 r:{\dotradius} \qbaresegment{#2}{#1}}

\newcommand{\qcross}[1]{\qcalcpos{#1} \move({\thori} {\vert})
                        \realmult{2.0}{\ballradius} \doubleballradius
                        \rlvec({\ballradius} {\ballradius})
                        \rmove(0 {-\doubleballradius})
                        \rlvec({-\doubleballradius} {\doubleballradius})
                        \rmove(0 {-\doubleballradius})
                        \rlvec({\doubleballradius} {\doubleballradius})}
\newcommand{\qsingle}[2]{\qbox{#1}{#1} \textref h:C v:C \htext{#2} \qskip}
\newcommand{\qmultiple}[3]{\qbox{#1}{#2} \textref h:C v:C \htext{#3} \qskip}
\newcommand{\qmeasure}[1]{\qcalcpos{#1} \move({\thori} {\vert})
                          \linewd {\dotradius} \lvec({\ttot} {\vert})
                          \linewd {\qlinewidth} \qbox{#1}{#1}
                          \textref h:C v:C \rmove(0 {-\halfministep})
                          \larc r:{\halfministep} sd:40 ed:140
                          \rmove(0 {\dotradius}) \arrowheadtype t:F 
                          \arrowheadsize l:{\dotradius} w:{\dotradius}
                          \realadd{\halfministep}{\dotradius}\arrowheight
                          \ravec({\dotradius} {\arrowheight})
                          \move({\thori} {\vert}) \qskip }
\newcommand{\qskipping}[2]{\qmultiple{#1}{#2}{\shortstack{.\\.\\.}}}
\newcommand{\qmulticontrol}[4]{\qsegment{#2}{#3} \qbox{#1}{#2}
                               \textref h:C v:C \htext{#4} \qskip}
\newcommand{\qcontrol}[3]{\qmulticontrol{#1}{#1}{#2}{#3}}
\newcommand{\qtwocontrol}[4]{\qsegment{#2}{#3} \qcontrol{#1}{#2}{#4}}
\newcommand{\qhadamard}[1]{\qsingle{#1}{H}}
\newcommand{\qrotcircle}[2]{\qcalcpos{#1} \move({\thori} {\vert})
                            \linewd #2 \fcir f:{\boxfill} r:{\halfministep}
                            \lcir r:{\halfministep} \linewd {\qlinewidth} }
\newcommand{\qrot}[2]{\qrotcircle{#1}{.5} \textref h:C v:C
                      \htext{\ensuremath{#2}} \qskip}

\newcommand{\qnot}[2]{\qsegment{#1}{#2} \lcir r:{\ballradius} 
                      \rlvec(0 {\ballradius}) \rlvec(0 {-\ballradius})
                      \rlvec(0 {-\ballradius}) \qskip}
\newcommand{\qswap}[2]{\qbaresegment{#1}{#2} \qcross{#1} \qcross{#2} \qskip}
\newcommand{\qcondrot}[3]{\qsegment{#1}{#2} \qrot{#1}{#3}}

\newcommand{\qtoffoli}[3]{\qsegment{#2}{#3} \qnot{#1}{#2}}
\newcommand{\qcalcbrace}[2]{\qcalcpos{#1} \realmult{\vert}{1}\bcenter
                            \qcalcpos{#2} \realadd{\vert}{\bcenter}\bcenter
                            \realadd{#2}{-#1}\bheight
                            \realadd{\bheight}{1}\bheight
                            \realmult{\bheight}{\step}  \bheight
                            \realmult{\bcenter}{.5}     \bcenter
                            \realmult{\bheight}{.5}     \bheight
                           }
\newcommand{\qbrace}[4]{\qcalcbrace{#1}{#2}
                        \move(0 {\bcenter}) \textref h:L v:C \htext{#3}
                        \move({\text} {\bcenter})
                        \qtextR{#4~\makebox[\ministep pt]%
                                 {$\left\{\rule{0pt}{\bheight pt}\right.$}}}
\newcommand{\qendbrace}[3]{\qcalcbrace{#1}{#2} \move({\ttot} {\bcenter})
                           \qtextL{\makebox[\ministep pt]%
                                {$\left.\rule{0pt}{\bheight pt}\right\}$}~#3}}
\newcommand{\qsidebar}[1]{\ifthenelse{#1 = 0}{\rmove({\halfministep} 0)}
                                             {\rmove({-\halfministep} 0)}
                          \rmove(0 {\halfmaxistep}) \linewd 3
                          \rlvec(0 {-\maxistep})
                          \ifthenelse{#1 = 0}{\rmove({-\halfministep} 0)}
                                             {\rmove({\halfministep} 0)}
                          \rmove(0 {\halfmaxistep}) \linewd {\qlinewidth}}

\begin{document}
\title{\huge \bf Toward an architecture\\for quantum programming}
\author{
  S.Bettelli\footnote{
    Laboratoire de Physique Quantique,
    Universit\'e Paul Sabatier,
    118, Route de Narbonne,
    31062 Cedex Toulouse (France),
    Tel.: \mbox{+33 05 61 55 65 73}
    email: bettelli@irsamc.ups-tlse.fr.},
  T.Calarco\footnote{
    National Institute of Standards and Technology,
    100 Bureau Drive, Stop 8423,
    Gaithersburg, MD 20899-8423 (USA),
    Tel.: \mbox{(301) 975-5347},
    Fax: \mbox{(301) 990-1350},
    email: Tommaso.Calarco@nist.gov.
    Also at: ECT*,
    European Centre for Theoretical Studies
    in Nuclear Physics and Related Areas,
    Villa Tambosi, Strada delle Tabarelle 286,
    38050 Villazzano (Italy),
    Tel.: \mbox{+39 0461 314738} (Room I/04),
    Fax: \mbox{+39 0461 935007},
    email: calarco@ect.it.}
  L.Serafini\footnote{
    Istituto Trentino di Cultura,
    Centro per la Ricerca Scientifica
    e Tecnologica (ITC-IRST),
    Via Sommarive 18 - Loc. Pant\`e,
    38050 Povo (Italy),
    Tel.: \mbox{+39 461 314319},
    Fax: \mbox{+39 461 302040},
    email: serafini@itc.it.}
}
\date{\today}

\maketitle

\abstract{
  It is becoming increasingly clear that, if a useful device for
  quantum computation will ever be built, it will be embodied by
  a classical computing machine with control over a truly quantum
  subsystem, this apparatus performing a mixture of classical and
  quantum computation. \par
  This paper investigates a possible approach to the problem
  of programming such machines: a template high level quantum
  language is presented which complements a generic general
  purpose classical language with a set of quantum primitives.
  The underlying scheme involves a run-time environment which
  calculates the byte-code for the quantum operations and pipes
  it to a quantum device controller or to a simulator. \par
  This language can compactly express existing quantum algorithms
  and reduce them to sequences of elementary operations; it also
  easily lends itself to automatic, hardware independent, circuit
  simplification. A publicly available preliminary implementation
  of the proposed ideas has been realized using the \cpp language.
}

\section{Quantum programming}
\label{sec:qprogramming}

\subsection{Introduction and previous results}

In the last decade the field of quantum computing has raised
large interest among physicists, mathematicians and computer
scientists due to the possibility of solving at least some
``hard'' problems exponentially faster than with the familiar
classical computers \cite{textbook}.
Relevant efforts have been concentrated in two directions:
on one hand a (still not so large) set of quantum algorithms
exploiting features inherent to the basic postulates of quantum
mechanics has been developed \cite{QAR}; on the other hand a
number of experimental schemes have been proposed which could
support the execution of these algorithms, moving quantum computation
from the realm of speculation to reality (see a review of basic
requirements in DiVincenzo \cite{checklist}). \par
The link between these two areas is a framework for describing
feasible quantum algorithms, namely a computational model, which
appears to have settled down to the quantum circuit model%
\footnote{Different computational models, involving physical
  systems with continuous-variable quantum systems
  used as computational spaces, are far less developed
  and will not be considered in this paper. It should
  be noted however that any such quantum device will
  most likely require very different interface abstractions
  from quantum circuits.},
due to Deutsch \cite{qcircuits-Deutsch}, Bernstein and Vazirani
\cite{qcircuits-Vazirani} and Yao \cite{qcircuits-Yao}. Though this
is satisfactory from the point of view of computational complexity
theory, it is not enough for a practical use (programming) of quantum
computers, once they will become available. \par
A few papers can be found in literature concerning the problem of 
{\em scalable} quantum programming. An unpublished report by Knill
\cite{knill}, gathering common wisdom of the period about the \qram
model (see section \ref{sec:qram_model}), moved the first steps
towards a standardised notation for quantum pseudo-code pointing
out some basic features of a quantum programming language (as an
extension of a conventional classical language), though the interest
was focused mainly on quantum registers (see section \ref{sec:regs}).
This report however did not propose any scheme for implementing an
{\em automatic} translation of the high level notation into
circuit objects. \par
Sanders and Zuliani \cite{zuliani_programming} extended the
probabilistic version of an imperative language (\pgcl) to include
three high level quantum primitives (initialisation, evolution and
finalisation). The resulting language (\qgcl) is expressive enough
to program a universal quantum computer, though its aim is more to
be a tool for verification of the procedures against their specifications%
\footnote{This work was extended in Zuliani's DPhil Thesis (Oxford
  University) submitted in July 2001, which however was not available
  at the time of writing.}
(i.e. for verifying that a program really implements the desired
algorithm) than to be the starting point for translation of quantum
specifications to low level primitives. A related paper by Zuliani
\cite{zuliani_pseudoclassical} showed an interesting technique for
transforming a generic \pgcl program into an equivalent but reversible
one, which has a direct application to the problem of implementing
quantum oracles for classical functions (see section
\ref{sec:pseudoclassical}). \par
B.\"Omer \cite{QCL1, QCL2} developed a procedural formalism (\qcl)
which shares many common points with the approach presented in this
article about the treatment of quantum registers%
\footnote{Quantum registers in \qcl are however dealt with in a non
  uniform fashion: in addition to \texttt{qureg}, two other register
  types are present, the \texttt{quvoid} and the \texttt{quscratch},
  which, for a proper type checking, require the knowledge of the
  quantum device state.}.
\qcl is however an interpreted environment and is not built on the
top of a standard classical language. In this language, just like in
\qgcl, non trivial unitary operations are functions (\texttt{qufunct}
or \texttt{operator}) instead of objects (see sections \ref{sec:qops}
and \ref{sec:opcomp}), so that their manipulation is subject to the
function call syntax; hence automatic operator construction (e.g.
controlled operators) and simplification are very difficult to
implement, if not impossible. Last, no notion of parallelism for
independent operators is present (see \ref{sec:lowlevel}). \par
In the following section, building on top of these previous works,
a list of desirable features for a quantum programming language
is presented.

\section{Desiderata for a quantum programming language}
A common theme in the field of quantum computation is the attempt
to think about algorithms in the new ``quantum way'', without being
misled by classical intuition. It could seem that describing quantum
computer algorithms in an almost entirely classical way would hide
rather than emphasise the difference between quantum and classical
computing. The point of this common objection is not, of course,
about criticising the assumption that the control system which drives
the evolution of the quantum device does not behave according to
classical mechanics. Rather, it could be originated by the guess
that regarding a part of the quantum resources as {\em program} and
another one as {\em data}, a sort of quantum von Neumann machine,
could lead more naturally to quantum algorithms. This guess has
however been disproved by Chuang and Nielsen \cite{gatearray},
who showed%
\footnote{The authors of \cite{gatearray} showed that a programmable
  quantum gate array, i.e. a quantum machine with a fixed evolution
  $G$ which deterministically implements the transformation
  $\ket{d}\otimes\ket{P_U} \rightarrow U\ket{d} \otimes \ket{P'_U}$,
  is such that if $U_1$ and $U_2$ are distinct unitary evolutions
  up to global phase changes, then \ket{P_{U_1}} and \ket{P_{U_2}}
  are orthogonal.
  \ket{P_U} here plays the role of the ``program'' and determines
  which operation $U$ is to be executed on the ``data'' register
  prepared in the state \ket{d}. The orthogonality of program states
  means that the program specification is indeed a classical 
  specification.}
that, if the program is to be executed deterministically, nothing
can be gained by specifying it through a quantum state instead of
through a classical one. These considerations can be summarised by
saying that quantum algorithms are specified inherently by means of
classical programming. \par
Subject of this article is the investigation and specification
of the desirable features of a quantum programming language.
The following list is a summary of the main points:
\begin{description}
\item[Completeness:]
  the language must be powerful enough to express the quantum
  circuit model. This means that it must be possible to code
  every valid quantum algorithm and, conversely, every piece
  of code must correspond to a valid quantum algorithm.
\item[Classical extension:]
  the language must include (i.e. be an extension of) a high
  level {\em classical computing paradigm} in order to integrate
  quantum computing and classical pre- and post-processing with the
  smallest effort. Ad hoc languages, with a limited implementation
  of classical primitives and facilities, would inevitably fall
  behind whenever ``standard'' programming technologies improve.
\item[Separability:]
  the language must keep classical programming and quantum
  programming separated, in order to be able to move to a
  classical machine all those computations which do not need,
  or which do not enjoy any speedup in being executed on,
  a quantum device.
\item[Expressivity:]
  the language must provide a set of high level constructs which
  make the process of coding quantum algorithms closer to the
  programmer's way of thinking and to the pseudo-code modular
  notation of current research articles. The language must allow
  an automated scalable procedure for translating and optionally
  optimising the high level code down to a sequence of low level
  control instructions for quantum machines. 
\item[Hardware independence:]
  the language must be independent from the actual hardware
  implementation of the quantum device which is going to be
  exploited. This allows ``recompilation'' of the code
  for different quantum architectures without the programmer's
  intervention.
\end{description}
The next sections are organised along the following lines.
After a general introduction about the computational model
(\ref{sec:qram_model}), the guidelines for the envisioned
language (\ref{sec:implementation}) are presented, together
with a discussion on hardware requirements (\ref{sec:hardware}).
Section \ref{sec:primitives} then describes the syntax for
high level constructs (\ref{sec:regs}, \ref{sec:qops}) as well
as the low level, but still hardware independent, primitives
to which these constructs get reduced (\ref{sec:lowlevel}).
Section \ref{sec:code} shows some code samples to clarify
the language layout. Section \ref{sec:internals} discusses
with more details some of the choices for the operator syntax
(\ref{sec:opcomp}) and the open problem of the implementation
of operators for classical functions (\ref{sec:pseudoclassical}).
The appendixes (\ref{app:implementation}, \ref{app:controlled},
\ref{app:translation}) present possible approaches for
implementing the high level primitives previously described.

\subsection{The \qram model}
\label{sec:qram_model}

\begin{figure}[b!]
  \begin{center}
    \includegraphics[width=.8\linewidth]{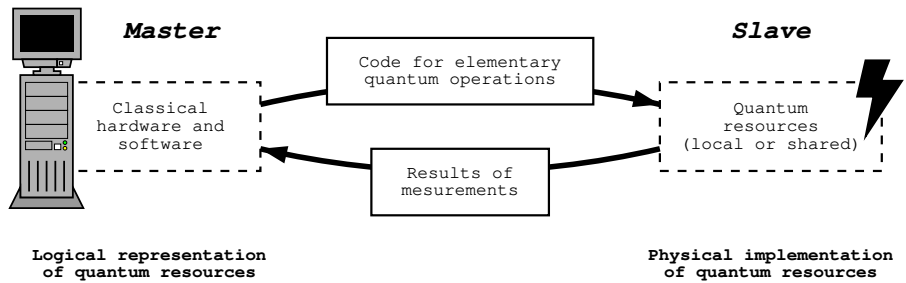}
    \caption{Simplified scheme of a \qram machine. The classical
      hardware drives the quantum resources in a master-slave
      configuration; it also performs pre-processing and
      post-processing of quantum data. The only feedback
      from the quantum subsystem is the result of measurements.}
    \label{fig:overview}
  \end{center}
\end{figure}

Before describing the structure of the proposed quantum language,
the quantum computer architecture which it is based on must be
clarified. Quantum algorithms are currently described by (more or
less implicitly) resorting to the \qram model (see e.g. Knill
\cite{knill}, Knill and Nielsen \cite{encMath}). \par
A \qram machine is an extension of a classical random access machine
which can exploit quantum resources and which is capable of all
kinds of purely classical computations. This classical machine
plays two roles: it both performs pre-processing and post-processing
of data for the quantum algorithms (trying to keep the quantum
processing part as limited in time as possible in order to help
preventing decoherence), and controls the quantum subsystem by
``setting'' the Hamiltonian which generates the required unitary
evolution, performing initialisations and collecting the results
of measurements.
In this scheme the quantum subsystem plays a slave role, while the
master classical machine uses it as a black-box co-processing unit.
This is summarised in the diagram in fig.\ref{fig:overview}. \par
It must be noted that quantum resources are not necessarily local%
\footnote{Knill \cite{knill} notes that situations arising in quantum
  communication schemes ``require operating on quantum registers
  in states prepared by another source (for example a quantum
  channel, or a quantum transmission overheard by an
  eavesdropper)''. It is likely however that these communication
  schemes will require quite different hardware, so that one
  would end up with two quantum subsystems better than with
  one but more complicated.};
they can be shared among different \qram machines for quantum type
communication or quantum distributed computing. This can be handled
by the \qram model if the machine is given access to the
hetero-controlled subsystem and to a classical synchronisation
system (a quantum network interface), but this article will not
delve into the details of these situations further. \par
The quantum resource, independently from its actual hardware
implementation, is treated as a collection of identical elementary
units termed {\em qubits}; a qubit is an abstract quantum system
whose state space is the set of the normalised vectors of the two
dimensional Hilbert space $\mathbb{C}^{\,2}$, and can encode as much
information as a point on the surface of a unit sphere (the Bloch
sphere). Due to the structure of the quantum state space, a qubit
can encode a superposition of the two boolean digits and is thus
more powerful than a classical bit, though the state can not be read
out directly. A collection of identical qubits is not subject to
the fermion or boson statistics, since the state space of the qubits
is in general only a portion of the state space of the quantum
system carrying the qubits, to which the statistics applies. \par

\subsection{A scheme for a quantum programming language}
\label{sec:implementation}

As already said, in order to perform a quantum computation, the
classical core of the \qram machine must modify the state of the
elements of the quantum subsystem it controls. In the proposed
language these elements are indexed by addresses. Though in the
following these addresses are treated like unsigned integer numbers,
thus abstracting the underlying quantum device to a linear structure,
it is by no means assumed that quantum memory is physically
organised as an array%
\footnote{It was so in very early schemes, like the seminal
  linear ion trap by Cirac and Zoller \cite{iontrap}, but
  many recent proposals with a concern to scalability are
  geared toward an at least two dimensional implementation.}.
The goal of the addresses for quantum elements is simply to hide to
the programmer the details of the memory handling. \par
It is well known that the {\em no-cloning} theorem excludes the
possibility of replicating the state of a generic quantum system%
\footnote{In other words, the transformation $\ket{\phi}\ket{0}
  \rightarrow \ket{\phi}\ket{\phi}$, where $\ket{0}$ is some
  fixed state and \ket{\phi} is variable, is prohibited in
  quantum mechanics, because it is not linear.}.
Since the call-by-value paradigm is based on the {\em copy}
primitive, this means that quantum programming can not use
call-by-value; therefore a mechanism for addressing parts of
already allocated quantum data must be supplied by the language. \par
\begin{figure}[!t]
  \begin{center}
    \includegraphics[width=.8\linewidth]{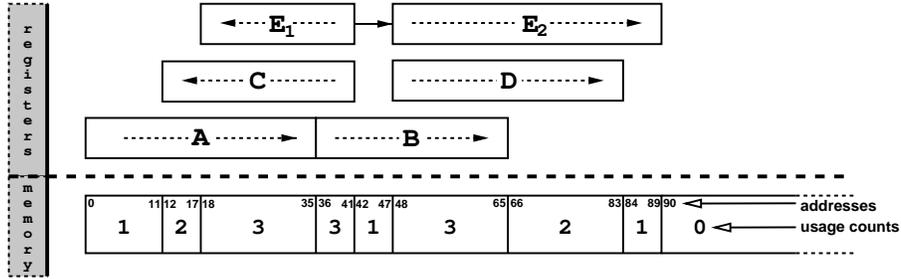}
    \caption{Examples of the organisation of quantum memory within
      registers. Each register is basically an ordered list of
      addresses with arbitrary size, like $A = [0, 35]$;
      registers may overlap, like $B$ and $D$ which both reference
      the $[48, 65]$ memory segment, or be ``inverted'',
      like $C = [41, 12]$. They can also be ``disjoint'',
      like $E = ([41, 18], [48, 89])$.
      The usage status of each qubit in the quantum memory array
      is maintained by the address manager, which is introduced
      in appendix \ref{app:implementation}. The qubits with
      addresses from $90$ on, in this example, are ``free''.}
    \label{fig:registers}
  \end{center}
\end{figure}

In view of these considerations, a new basic data type, the
{\em quantum register} is introduced in the proposed quantum
computing language. Quantum register objects are arbitrary
collections of distinct qubit addresses. Arbitrary means both
that the size is bounded only by the amount of available quantum
resources and that the addresses need not be contiguous.
Moreover, different quantum registers may overlap, so that the
same address can be contained in more than one collection;
an example of a set of overlapping registers is shown in
fig.\ref{fig:registers}. A qubit is ``free'' if its address
is not referenced by any existing quantum register. \par
The second data type which is proposed for the quantum language
is the {\em quantum operator}. A quantum operator object encodes
the definition of a generic quantum circuit (an acyclic network
of quantum gates) and, when fed with a quantum register, it is
able to execute the circuit on the supplied register.
The operators are the interface which is presented to the
programmer for handling unitary transformations. The action of
a quantum operator object onto a quantum register object produces,
transparently to the programmer, a stream of {\em byte-code}
(a sequence of quantum gate codes and the addresses upon which
they must be executed) to be fed into the interface to the quantum
device. \par
Since a quantum operator object embeds the definition of the
corresponding circuit as a datum, it is possible to automatically
manipulate this definition in a number of different ways, for
instance for creating the composition of a pair of circuits,
building derived operators (like controlled or adjoint ones)
and running simplification routines. \par
A more detailed description of the proposed scheme is given in
appendix \ref{app:implementation}.

\subsection{Assumptions on quantum hardware}
\label{sec:hardware}

The actual construction of a quantum computer is a tremendous
challenge for both experimental and theoretical researchers in
the field. At the moment, it is simply unknown what kind of quantum
mechanical system is the most useful for this task.
As a result, it is also unknown what kind of {\em low level}
architecture is the most suitable for the implementation of a
quantum computer. This is an additional reason, beyond general
considerations on the design of high level programming languages,
in favour of not referencing a specific physical system that
is supposed to support the language design. \par
The real computational complexity of a quantum algorithm however
depends on which capabilities the quantum hardware is endowed with.
Though the programming language must be as general as possible, hence
adaptable to a variety of quantum devices, some minimal assumptions
on the \qram machine have been made in this paper. \par
The first assumption concerns the hardware implementation of
primitive quantum gates. It seems very plausible that single-qubit
gates as well as two-qubit gates between neighbouring quantum
subsystems can be executed in constant time, independently of the
physical location of the involved qubit(s); two-qubit gates between 
non-neighbouring locations on the other hand are more challenging.
In most of the currently proposed schemes these gates can be
implemented only locally, and physical qubit swapping is needed in
order to fulfil this condition; this means that the execution time
of a two-qubit gate scales linearly with the ``distance'' (the number
of required swaps) between the qubit locations. \par
Since the physical layout of a quantum device remains unknown to
a high level programming language, it is not possible to specify
in the language any closeness relation, and the hardware independent
language environment can not help but assume that two-qubit gates
get executed in constant time, and optimise its behaviour with
respect to this assumption. This means, in practice, that the real
``complexity'' of any specified circuit is in general implementation
dependent, and worse than that presented by the language. \par
The second assumption concerns the implementation of parallelisable
gates. A set of applied gates is parallelisable if every qubit in
the quantum device is non trivially affected by at most one gate.
A parallel quantum machine device is such if it is able to execute
a parallelisable set of gates in a time bounded by the execution
time of the most expensive gate. This capability is also required
in order to perform fault tolerant quantum computation%
\footnote{See the discussion in \cite{textbook}, sec.10.6.4 at
  pag.493 about the threshold theorem for quantum computation.}.
The language assumes that the underlying quantum hardware is
parallel; indeed, only the parallelisation of homogeneous gates
(see page \pageref{sec:qops:ctors}) is actually exploited.

\section{Language primitives}
\label{sec:primitives}

In this section the set of primitives for the proposed language
is introduced. The first part concerns the high level primitives
(HLP) for the construction and manipulation of quantum registers
(\ref{sec:regs}) and quantum operators (\ref{sec:qops}).
Register and operators are classical data structures; their handling
does not require any interaction with the quantum device, exception
made for the application of an operator to a register, and the
initialisation or measurement of a register. \par
The second part (\ref{sec:lowlevel}) introduces a set of low level
primitives (LLP) to which the high level specification of a quantum
program gets reduced, transparently to the programmer. The LLP are
the actual ``code'' which is sent to the generic quantum device.

\subsection{Register handling}
\label{sec:regs}

As explained in the introductory section, a consequence of the basic
rules of quantum mechanics is that an unknown generic state of a
quantum register can not be inquired without being destroyed. Since
such state encodes the intermediate steps of the computation, it must
de facto be regarded as unknown; hence quantum registers can not be
read while the computation is running on them.
The main implication is that the programmer must think of a quantum
register (\texttt{Qreg}) as an {\em interface} to a portion of the
quantum device, not as an object carrying a value, unless, of course,
he decides to measure it. The register is equivalent to a list of
distinct%
\footnote{Distinctness is required because multi-qubits operations
  need distinct physical locations; by assuming that all registers
  contain by construction only distinct addresses, checking this
  condition when a register is fed into a quantum operation object
  can be avoided.}
addresses in the quantum memory, and the language provides a set
of compliant operations on such lists, which are specified in the
following. While all addresses in the same register are distinct, the
impossibility to make copies of the register content requires the
ability to have more than one register reference the same
addresses. \par
The initialisation and measurement primitives for registers are
described in section \ref{sec:lowlevel}, since they are not distinct
from the corresponding LLP. All those primitives which are listed
here do not involve any interaction with the quantum device. \par

\begin{myenum}
  
  \myitem{Register allocation} \label{sec:regs:alloc}
  Register allocation is the action of creating a quantum register
  which references only free qubit locations (i.e. not already
  referenced by an existing register). Quantum registers can be
  created with arbitrary size, limited only by the capacity of the
  quantum device, the smallest register interfacing to a single
  qubit. It is possible to inquire the size of a register. More
  details on how the allocation status of qubit locations can be
  handled are given in appendix \ref{app:implementation}. \par
  \myex{Qreg a\_register(5);}
  {allocates a register with 5 qubits} \\
  \myex{int the\_size = a\_register.size();}
  {inquires the size of the register}

  \myitem{Register addressing and concatenation} \label{sec:regs:addconc}
  As already explained, the programmer must be able to operate
  on registers which overlap parts of already existing ones.
  Therefore the register objects support the addressing operation
  (the creation of a register from a subsection of an old register)
  and the concatenation operation (the creation of a register from
  the juxtaposition%
  \footnote{Concatenation requires more bookkeeping than simple
    addressing, because it must be checked that all the addresses
    in the composed register are distinct.}
  of two old registers). A proper combination of these two
  operations allows for the creation of a register which is the
  most general reorganisation of the used portion of the quantum
  device. \par
  \myex{Qreg a\_qubit = a\_register[3];}
  {selects the fourth qubit from \texttt{a\_register}} \\
  \myex{Qreg a\_subreg = a\_register(2,5);}
  {selects 5 qubits starting at the third one} \\
  \myex{Qreg new\_reg = a\_subreg \& a\_qubit;}
  {concatenates the two registers}
  
  \myitem{Register resizing} \label{sec:regs:resize}
  Once a register object has been created, it can be resized
  (extended or reduced) by adding new qubits or dropping some of
  them at its beginning. This ability is very useful for routines
  which need to spawn and reabsorb auxiliary qubits during their
  execution, provided they take care of a proper uncomputation.
  Extending a register works like spawning a new register with the
  required additional size, concatenating it with the old register
  and renaming the latter. Dropping qubits works like the
  deallocation of only a part of it, taking care that the reduced
  register contains at least one address. \par
  \myex{my\_register += 5;}
  {adds five qubits to \texttt{my\_register}} \\
  \myex{my\_register -= 3;}
  {drops three qubits from \texttt{my\_register}}

  \myitem{Register deallocation} \label{sec:regs:dealloc}
  Register deallocation is the act of destroying the classical
  object which represents the interface to a portion of the
  quantum device. Before being eliminated, this object must
  release the allocated resources. As a consequence, the ``usage''
  of a part of the quantum device can drop to zero, which means
  that that part is free for a new allocation. \par

\end{myenum}

\subsection{Quantum Operators and their manipulation}
\label{sec:qops}

Quantum operator objects (\texttt{Qop}) are the counterpart in the
proposed language of quantum circuits, that is unitary transformations
on the finite dimensional Hilbert space of a register ($U(2^n)$ if $n$
is the register size). The action of quantum operators is to modify
the state of a part of the quantum device, interfaced by a
register. \par
As it is well known%
\footnote{See \cite{textbook}, sec.4.5.1 and 4.5.2 at pag.189.},
all such unitary transformations can be built by finite composition
using only matrices acting non-trivially on a one- or two-level
subsystem of the original Hilbert space; their number is in general
exponential in $n$. Furthermore, it is possible to approximate%
\footnote{The approximation of a single-qubit gate is very efficient:
  the Solovay-Kitaev theorem proves that an arbitrary single-qubit
  gate may be approximated to accuracy $\epsilon$ using
  $O(\log^c(1/\epsilon))$ gates from a discrete set, where $c\sim 2$.
  The approximation of two-qubit gates can be efficiently reduced
  to previous case if a non trivial two-qubit gate is available as
  a primitive. The problem of approximating a generic transformation
  is however very hard: there are unitary transformations on $m$
  qubits which take $\Omega(2^m\log(1/\epsilon)/\log(m))$ operations
  from a discrete set to approximate. See \cite{textbook}, sec.4.5.3
  and 4.5.4 at pag.194.}
each of these matrices using a finite gate subset%
\footnote{See \cite{textbook}, sec.4.5.3 at pag.194.}
containing some single-qubit operations and one two-qubit operation
(see section \ref{sec:lowlevel}). \par
The decomposition into these LLP has exponential complexity in
general, but this is not the case, of course, for efficient quantum
algorithms, which have both time (circuit depth) and space (register
size) requirements which are polynomial in the input size; on the
other hand, representing a transformation by its matrix elements
without any compression scheme is always exponential in the input
size. An efficient scheme for quantum operators should therefore be
{\em (de)composition oriented}, i.e. an operator should be stored
as the sequence of its factors. \par
The following list describes the HLP which can be used by the
programmer in order to specify a quantum operator. The construction
of a quantum operator is a purely classical computation, it does
not need to reference quantum registers and must use a polynomial
amount of classical resources (space for storage and time for
calculations) in order to be useful.

\begin{myenum}

  \myitem{Identity operator}
  A quantum operator object can be constructed without any parameter;
  in this case it corresponds to the identity operator over a quantum
  register with arbitrary size. It can then be extended using operator
  composition, see point \ref{sec:qops:comp}. \par
  \myex{Qop my\_op;}
  {constructs the identity operator}
  
  \myitem{Fixed arity quantum operators} \label{sec:qops:ctors}
  Each primitive fixed arity quantum operator is associated to a
  matrix $M$ acting on $k$ qubit lines, i.e. a matrix in $U(2^k)$,
  and is specified by $k$ index lists $\{\ell^{\;(h)}\}_{h\in [0,k[}$
  (all the lists have the same size $s$, and all the indexes are
  distinct even among different lists). The action of such an
  operator onto a register is to apply $M$ to the qubits in the
  register indexed by $\ell^{\;(0)}_j$, \dots, $\ell^{\;(k-1)}_j$
  for each $j\in[0,s[$. In simpler words, a single primitive
  fixed arity quantum operator represents a circuit with $s$ copies%
  \footnote{Such a way of representing quantum operations reduces the
    amount of classical resources needed to store the quantum circuit
    and becomes very useful when the quantum device is able to run
    a number of independent copies of a quantum gate in parallel.}
  of the matrix $M$ in parallel. \par 
  \picinpar
  {The order of indexes inside a list is only relevant with respect
    to the order of indexes in the first list (hence the order of the
    single list of a single-qubit primitive is arbitrary). An example
    of how these lists are used is shown in the picture on the right:
    the circuit corresponds to the creation of a CNOT operator with
    control index list $\ell^{\;(0)}\!\!= \!(0,4,5)$ and target index
    list $\ell^{\;(1)}\!\!=\!(1,2,6)$. The symbol used for a CNOT
    gate is \raisebox{-2pt}[0pt][0pt]
    {\begin{qcircuit}{6}{1}{2}{0}\qnot{0}{1}\end{qcircuit}}. \par
    A summary of primitive quantum operators can be found in table
    \ref{tab:cprim}.}
  {\begin{qcircuit}{12}{1}{7}{10}
      \qnot{5}{6} \qsame \qnot{4}{2} \qsame \qnot{0}{1} {\tiny
        \qlabel{0}{6} \qlabel{1}{5} \qlabel{2}{4}
        \qlabel{3}{3} \qlabel{4}{2} \qlabel{5}{1} \qlabel{6}{0} }
    \end{qcircuit}}
  \par
  \myex{Qop my\_op = QHadamard(7);}
  {Hadamard gates acting on first 7 qubits} \\
  \myex{Qop my\_op = QCnot(ctrls, targets);}
  {see above (\texttt{ctrls} $=\ell^{\;(0)}$,
    \texttt{targets} $=\ell^{\;(1)}$)}
  
  \myitem{Macro quantum operators} \label{sec:qops:macro}
  The primitive quantum operators previously described correspond
  to fixed arity quantum gates applied in a parallel fashion. It
  is useful to consider also a different type of high level primitive
  (a macro from now on) which is associated to a single transformation
  in $U(2^n)$ homogeneously parametrised with respect to the number
  $n$ of qubit lines. A macro with a given dimension $n$ can not in
  general be reduced to a tensor product of macros of the same type
  with a smaller dimension. A macro is defined by a single list of
  addresses, whose order is meaningful (differently from fixed
  arity quantum operators with arity equal to one). \par
  Quantum macros could become very handy if a specific quantum hardware 
  is built which is able to implement the macro more efficiently 
  than by running the corresponding sequence of less specific low
  level primitives. In general however their constructor expands the
  macro into an equivalent sequence of fixed arity operators. \par
  \myex{Qop my\_op = QFourier(7);}
  {Fourier transform on the first 7 qubits}

  \myitem{Qubit line reordering} \label{sec:qops:reorder}
  Some quantum operations%
  \footnote{the best known example is the quantum Fourier transform,
    see \cite{QFT_orig}, where the least significant qubits of the 
    input are transformed into the most significant qubits of the
    output and vice versa.}
  require a permutation of the qubits inside the register they
  operate on, for instance in order to preserve a standard
  convention for the most or least significant location.
  This can be accomplished by properly exchanging the quantum
  states of the qubits referenced by the register. The language
  provides a fixed arity primitive (with arity equal to two)
  which performs such exchanges. \par
  Running this swap operation on the quantum device is however
  a waste of computation time since it is a completely 
  {\em classical} data manipulation. Appendix \ref{app:translation}
  describes a possible approach for an implementation which,
  transparently to the programmer, reorganises (on the classical
  machine) the mapping between qubit addresses and qubit locations,
  achieving the same result. \par
  \myex{Qop a\_swap = QSwap(5);}
  {implements the swap of the first 5 qubits}

  \myitem{Controlled operators} \label{sec:qops:ctrl}
  A controlled-$U$ operator is a quantum operator $C_U$ which
  implements the transformation
  $C_U\ket{x}\ket{y} = \ket{x}U^{\delta_{x,1\dots1}}\ket{y}$,
  that is it applies $U$ to the second register when the first is
  found in the state \ket{1\dots1}. It is a very useful high level
  primitive for quantum algorithms. This operator is of course
  unitary and its adjoint is the $C_{U^\dagger}$ operator.
  Quantum operators have a constructor for such controlled 
  objects, taking as input the operator to be controlled and the
  size of the control register. They need, in general, to use
  ancilla qubits during their execution, which are to be supplied
  by the language internals transparently to the user.
  Some techniques for the implementation of controlled operators
  are discussed in appendix \ref{app:controlled}. \par
  \myex{Qop a\_controlled\_op(U, 5);}
  {creates a $U$ conditioned by 5 qubits}

  \myitem{Operators for classical functions} \label{sec:qops:classical}
  Given an algorithm for a classical function $f : \mathbb{Z}_{2^n}
  \rightarrow \mathbb{Z}_{2^m}$, it is often of interest in quantum
  computation the mapping $U_f\ket{x}\ket{y} = \ket{x}\ket{y\oplus
    f(x)}$ where $\oplus$ is the bitwise \texttt{XOR}, the two
  registers having respectively size $n$ and $m$. These operators,
  which implement a classical function in a reversible fashion, are
  always self-adjoint%
  \footnote{Since $U_f^2 \ket{x}\ket{y} = U_f \ket{x}\ket{y\oplus f(x)}
    = \ket{x}\ket{y\oplus f(x)\oplus f(x)} = \ket{x}\ket{y}$},
  generally create entanglement between the input and output registers
  and are necessary to insert non-injective classical functions in the
  quantum computing scheme%
  \footnote{Implementation of classical functions is for instance
    needed in the Grover's algorithms \cite{grover}, where they are
    used to evaluate the fitness of a candidate solution to an $NP$
    problem.}. \par
  A quantum language needs the ability to  build the $U_f$ operator
  automatically once the programmer has specified an algorithm for
  $f$ using the formalism of the underlying classical language.
  If $f$ is boolean (that is $m=1$), an easy construction with an
  additional ancilla qubit can implement the ``phase'' mapping
  $P_f \ket{x} = (-1)^{f(x)}\ket{x}$. For a longer discussion about
  the problems this facility rises refer to section
  \ref{sec:pseudoclassical}. \par
  \myex{Qop an\_oracle = Qop(f,3,5);}
  {oracle for $f$ with $n=3$ and $m=5$} \\
  \myex{Qop a\_phase\_oracle = Qop(g,4);}
  {phase oracle for $g$ with $n=4$ ($m$ is 1)}
  
  \myitem{Operator composition} \label{sec:qops:comp}
  Composing two quantum operator objects returns an operator
  which represents the concatenation of the underlying circuits
  in the specified order (i.e. the first operator gets executed
  first, similarly to how circuits are drawn and differently from
  the mathematical notation, where operators act on the right).
  A more elaborated analysis of the advantages of a composition
  oriented representation can be found in section
  \ref{sec:opcomp}.\par
  The language provides three different versions of the composition
  of operators, in order to achieve increasing efficiency by
  reusing existing data structures: {\em concatenation} leaves
  the two argument operators untouched and returns a third object,
  {\em augmentation} modifies the first argument to hold the 
  composed operator without modifying the second argument and
  {\em splicing} moves all the data from the second operator (which
  is left the identity) into the first. \par
  \myex{Qop composed = part\_1 \& part\_2;}
  {composes two \texttt{Qop}s into a third \texttt{Qop}} \\
  \myex{my\_operator \&= an\_operator;}
  {extends \texttt{my\_operator} with \texttt{an\_operator}} \\
  \myex{my\_operator << an\_operator;}
  {moves \texttt{an\_operator} into \texttt{my\_operator}}
  
  \myitem{Operator conjugation} \label{sec:qops:conj}
  Given a quantum operator $U$, it is possible to specify its
  adjoint $U^{\dagger}$ with the {\em conjugation} transformation.
  Conjugation may act on the quantum operator object in place or
  create a new \texttt{Qop}. The proposed quantum language supplies
  both the mutating and the non mutating transformations. \par
  \myex{Qop adj\_operator = !an\_op;}
  {creates the adjoint operator} \\
  \myex{an\_op.adjoin();}
  {conjugates the operator}

  \myitem{Operator permutations} \label{sec:qops:perm}
  Quantum operators need a method for rearranging the order of
  the indexes of qubit lines for the underlying circuits, so that
  simpler operators can be adapted to fit as modules into more complex
  ones; an example of this is described in section \ref{sec:3adder}
  where two copies of a circuit for performing the addition of two
  input registers are rearranged to build a circuit for the addition
  of three input registers. \par
  A generic permutation can be decomposed into a sequence of adjacent
  transpositions, but this approach would be highly inefficient in
  most situations; it is better to have access to ``higher level''
  permutations which can modify the index lists in a single step.
  The proposed language supplies the {\em split} and {\em invert}
  permutations, both in their mutating and non mutating version.
  A split permutation leaves the first part of the circuit unaltered
  while ``shifting down'' the rest of it; an invert permutation instead
  ``reverses'' the central part of a circuit. Shifting the whole
  circuit ({\em offset }) is a sub-case of the split operation.
  A better understanding of these two manipulators can be gained
  visually with figure \ref{fig:permutations}. \par
  \myex{Qop split = an\_op(2,3,SPLIT);}
  {creates a split operator} \\
  \myex{Qop inverted = an\_op(2,3,INVERT);}
  {creates an inverted operator} \\
  \myex{Qop shifted = an\_op >> 2;}
  {creates an offset operator} \\
  \myex{an\_op.offset(2)\hspace{-2pt}.invert(2,3)\hspace{-2pt}.split(2,3);%
    \hspace{-10pt}}
  {offsets, inverts and splits the operator}
  
  \begin{figure}[tb!]
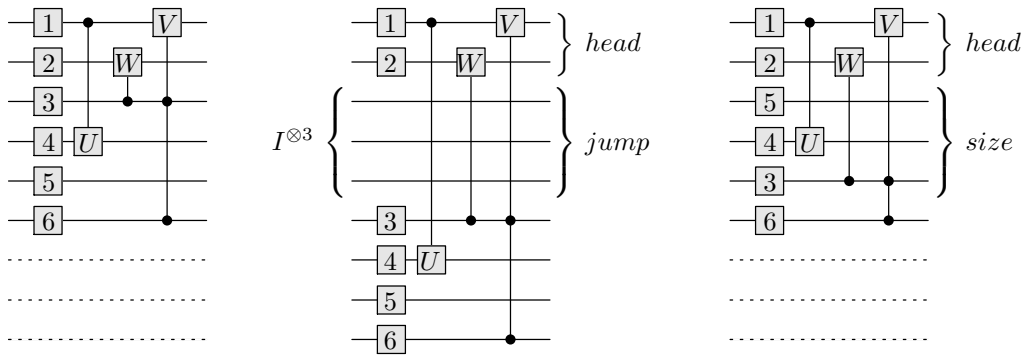

    \hfill
    \begin{qcircuit}{15}{4}{9}{0}
      \qpatchline{0}{0}{5}{1 3} \qpatchline{1}{0}{5}{1 3}
      \qpatchline{2}{0}{5}{1 3}
      \qsingle{3}{6} \qsame \qsingle{4}{5} \qsame \qsingle{5}{4} \qsame
      \qsingle{6}{3} \qsame \qsingle{7}{2} \qsame \qsingle{8}{1}
      \qcontrol{5}{8}{$U$} \qcontrol{7}{6}{$W$} \qtwocontrol{8}{6}{3}{$V$}
    \end{qcircuit}
    \hfill
    \begin{qcircuit}{15}{4}{9}{0}
      \qsingle{0}{6} \qsame \qsingle{1}{5} \qsame
      \qsingle{2}{4} \qsame \qsingle{3}{3} \qsame
      \qsingle{7}{2} \qsame \qsingle{8}{1}
      \qcontrol{2}{8}{$U$} \qcontrol{7}{3}{$W$} \qtwocontrol{8}{3}{0}{$V$}
      \qbrace{4}{6}{}{$I^{\otimes 3}$}
      \qendbrace{7}{8}{$head$} \qendbrace{4}{6}{$jump$}
    \end{qcircuit}
    \hfill
    \begin{qcircuit}{15}{4}{9}{0}
      \qpatchline{0}{0}{5}{1 3} \qpatchline{1}{0}{5}{1 3}
      \qpatchline{2}{0}{5}{1 3}
      \qsingle{3}{6} \qsame \qsingle{4}{3} \qsame
      \qsingle{5}{4} \qsame \qsingle{6}{5} \qsame
      \qsingle{7}{2} \qsame \qsingle{8}{1}
      \qcontrol{5}{8}{$U$} \qcontrol{7}{4}{$W$} \qtwocontrol{8}{4}{3}{$V$}
      \qendbrace{7}{8}{$head$} \qendbrace{4}{6}{$size$}
    \end{qcircuit}
    \hfill~
    \caption[Quantum operator permutations]{
      This figure illustrates the two kinds of index permutations for 
      quantum operators. \underline{Left}: a generic quantum circuit
      with various types of quantum gates. \underline{Centre}: the same
      circuit after a {\em split(head, jump)} operation with {\em head}
      set to 2 and {\em jump} set to 3; the first {\em head} qubit
      lines are left untouched, while the others are shifted down by
      {\em jump} lines. \underline{Right}: the same circuit after an
      {\em invert(head, size)} operation with {\em head} set to 2 and
      {\em size} set to 3; the first {\em head} qubit lines are left
      untouched, the following {\em size} lines are inverted and all
      the remaining circuit is unmodified.}
    \label{fig:permutations}
  \end{figure}
  
  \myitem{Application of an operator} \label{sec:qops:exec}
  Quantum operators must have a method for running the circuit they
  embed onto a quantum register supplied by the programmer. Executing
  an operator means executing all of its factors in sequence: the index
  lists in each primitive operator must be coupled with the address
  lists in the given quantum register in order to calculate the qubits
  to be addressed, and the appropriate byte-code must be sent to the
  quantum device%
  \footnote{If a quantum operator must be repeated on the same register
    a number of times, a mechanism could be provided for caching the 
    byte-code and resend it without recalculating all address pairings
    from the beginning each time.}.
  This process may require ancilla qubits, which need to be spawned
  and reabsorbed transparently to the user. See appendix
  \ref{app:translation} for more details on the steps taken when an
  operator is executed. \par
  \myex{an\_operator(a\_register);}
  {runs the circuit onto the register}

\end{myenum}

\subsection{Low level primitives}
\label{sec:lowlevel}

Low level primitives are the basic building blocks for the
communication between the language and the quantum device.
They are divided in non unitary (initialisation and measurement)
and unitary (quantum gates). Quantum gates are used to build up all
quantum circuits and must of course form a complete set (redundancy
is not a problem); choosing a universal set of gates together with
a proper syntax for them (see the previous section and \ref{sec:opcomp})
ensures that all and only quantum circuits in the \qram model can be
expressed by the proposed quantum language. \par
Initialisations and measurements, which are not unitary, are operated
directly onto quantum registers. Since registers can have arbitrary
sizes, the assigned or returned values do not in general fit into a
standard integer type of the classical language, therefore a new type
for ordered sets of bits should be introduced (\texttt{Qbitset} in
the following), with automatic conversion to/from unsigned integers
when possible. \par
As remarked in section \ref{sec:qops}, an efficient scheme for
quantum operators should store quantum circuits using one of their
factorisations. The smallest factors are called in the following
{\em time slices}. In the proposed language, in a way similar to
primitive quantum operators, each time slice is not simply a quantum
gate, but embeds a sort of parallelisation restricted to homogeneous
gates, which can be acted in parallel over multiple independent qubits.
The quantum programmer does not however deal directly with time slices,
but he uses only the set of high level primitives described in the
previous section. \par
Storing quantum primitives as a list of time slices fits nicely with
the previous requirements for quantum operators, e.g conjugation is
easily achieved by iterating through the list in the reverse order
and conjugating all its elements%
\footnote{All the quantum gates corresponding to time slices should
  have their adjoint gate implemented as a primitive, so that each
  quantum operator and its adjoint have exactly the same circuit
  depth.};
splicing (the third version of operator composition) requires
constant time.

\begin{myenum}
  
  \myitem{Register initialisation and assignment} \label{sec:lowlevel:assign}
  The most obvious primitive for a quantum register is its
  {\em initialisation} to an element of the computational basis. On
  a realistic quantum device this involves setting all the qubits of
  the register to some reference state (e.g. the ground state) and
  subsequently performing the required unitary transformation
  to turn it to the representation of an arbitrary integer. It is
  evident that {\em assignment} of a \texttt{Qbitset} to a quantum
  register is the same operation as before, and involves a
  re-preparation of qubits of the register. \par
  \myex{Qreg a\_register(5,3);}
  {initialises a 5 qubits register to \ket{3}}\\
  \myex{a\_register = 7;}
  {prepares the register again in \ket{7}}
  
  \myitem{Register measurement} \label{sec:lowlevel:measure}
  The programmer must be able to {\em measure} a register obtaining
  an element of the computational basis (that is an integer number
  or a sequence of boolean values) to be used in the following of
  the algorithm. This operation is the only blocking primitive with
  respect to the code flow in the classical core, because the classical
  environment must wait for the quantum device to execute all the
  generated byte-code, perform the measurement and return the
  result. \par
  \myex{Qbitset val = a\_register.measure();}
  {measures a register and saves the result}\\
  \myex{int val = a\_register.measure();}
  {casts to integer if possible}

  \myitem{Low level unitary gates} \label{sec:lowlevel:prim}
  As already said, it is not important which low level unitary gates
  are chosen to implement a version of the proposed quantum language,
  as long as the set is complete: this ability to switch to another
  set must be retained, since it is far from obvious which primitives
  will most easily be implemented and standardise in future quantum
  computers. \par
  In this paper (see table \ref{tab:cprim}) the Welsh-Hadamard
  transform $H$, the enumerable set of phase shifts $R_k$ (rotations
  around the $z$-axis) and their controlled counterparts $C_{R_k}$
  are used as a complete%
  \footnote{This set is redundantly universal; note that
    $(\mathbb{I}\otimes H) \circ C_{R_1} \circ (\mathbb{I}\otimes H)$
    is the CNOT gate, $R_2$ is the ``phase'' gate and $R_3$ is the
    ``$\pi / 8$'' gate. CNOT, ``phase'' and ``$\pi / 8$'', together
    with $H$ are the so called {\em standard set} of universal gates
    (see \cite{textbook} at pag.195).}
  set of unitary LLP.
  Let $\phi_k$ be $e^{2\pi i / 2^k}$ for $k \in \mathbb{N}$ and
  its conjugate $e^{- 2\pi i / 2^{|k|}}$ for $k \in \mathbb{Z}/
  \mathbb{N}\,$; then the matrix representation of these LLP is
  as follows:
  \begin{displaymath}
    H=\frac{1}{\sqrt{2}}\left(\begin{array}{cc}
        1 & 1 \\ 1 & -1 \end{array}\right)
    \begin{qcircuit}{15}{0}{0}{0}\qhadamard{0}\end{qcircuit}
    \quad \quad
    R_k=\left(\begin{array}{cc}
        1 & 0 \\ 0 & \phi_k \end{array}\right)
    \begin{qcircuit}{15}{0}{0}{0}\qrot{0}{k}\end{qcircuit}
    \quad \quad
    C_{R_k}=\left(\begin{array}{cc}
        \mathbb{I} & 0 \\ 0 & R_k \end{array}\right)
    \begin{qcircuit}{15}{0}{0}{0}\qcondrot{0}{1}{k}\end{qcircuit}
  \end{displaymath}
  Moreover, $H$ is self-adjoint and $R_k$ is the adjoint of $R_{-k}$
  ($C_{R_k}$ is the adjoint of $C_{R_{-k}}$), hence this set is closed
  under conjugation. In appendix \ref{app:controlled} it is shown that
  $C_U$, where $U$ is one of the previous gates, can be expanded into
  a circuit of gates from the same set with depth bounded by a constant.
  Primitive quantum operations are built using LLP, but they are
  logically distinct: there is no need for a one to one
  correspondence. \par
  \picinpar
  {This decoupling allows more portable quantum code to be written,
    since the translation from HLPs into LLPs can be delegated to
    ``more hardware-specific'' libraries. Expanding on a previous
    example, in the picture on the right it is shown the circuit
    corresponding to the creation of a CNOT operator with control
    indexes $(0,4,5)$ and target indexes $(1,2,6)$, reduced to LLP.
    The relevant identity is $X = H R_1 H$, where $X$ is the NOT port.}
  {\begin{qcircuit}{12}{1}{7}{10}
      \qnot{5}{6} \qsame \qnot{4}{2} \qsame \qnot{0}{1} {\tiny
        \qlabel{0}{6} \qlabel{1}{5} \qlabel{2}{4}
        \qlabel{3}{3} \qlabel{4}{2} \qlabel{5}{1} \qlabel{6}{0} }
    \end{qcircuit}
    $\equiv$
    \begin{qcircuit}{12}{3}{7}{0}
      \qhadamard{5} \qsame \qhadamard{4} \qsame \qhadamard{0}
      \qcondrot{5}{6}{1} \qsame \qcondrot{4}{2}{1} \qsame \qcondrot{0}{1}{1} 
      \qhadamard{5} \qsame \qhadamard{4} \qsame \qhadamard{0}
    \end{qcircuit}}
  
\end{myenum}

\section{Code fragments}
\label{sec:code}

A preliminary implementation of the ideas presented in the previous
sections has been developed in the form of a \cpp \cite{C++} library
by the authors and is freely available on the Internet. This section
introduces the flavour of the proposed high level language by showing
some examples of source code. The code is of course not optimised in
order to be more understandable. The \cpp like syntax is summarised
in tables \ref{tab:regs}, \ref{tab:qops} and \ref{tab:cprim}, but they
are not strictly necessary in order to follow the discussion.

\subsection{A three-input adder}
\label{sec:3adder}

This example illustrates how operator compositions, permutations and
adjoining can be used in the classical preprocessing stage in order
to build a complex parametric quantum operator by reusing smaller
circuits. \par
The following circuit implements the core of the quantum Fourier
transform \cite{QFT_orig} for a four-qubit register, where
\ket{\varphi(\alpha)} stands for $\frac{1}{\sqrt{2}} \left( \ket{0} +
  e^{2\pi i \alpha}\ket{1} \right)$. The circuit is different from
that usually reported on quantum computing textbooks since the final
rearrangement of qubit lines is not performed. For this reason the
corresponding unitary operator is named $\widetilde{\mathcal{F}}$. The
effect of the transformation is to move information from the
computational basis representation into the phase coefficients.

\begin{center}
  \begin{qcircuit}{12}{10}{4}{25}
    \qlabel{0}{\ket{x_3}} \qlabel{1}{\ket{x_2}}
    \qlabel{2}{\ket{x_1}} \qlabel{3}{\ket{x_0}}
    \qhadamard{3} \qcondrot{3}{2}{2} \qcondrot{3}{1}{3} \qcondrot{3}{0}{4}
    \qhadamard{2} \qcondrot{2}{1}{2} \qcondrot{2}{0}{3}
    \qhadamard{1} \qcondrot{1}{0}{2} \qhadamard{0}
    \qendlabel{0}{$\ket{\varphi(0.x_3)}$}
    \qendlabel{1}{$\ket{\varphi(0.x_2x_3)}$}
    \qendlabel{2}{$\ket{\varphi(0.x_1x_2x_3)}$}
    \qendlabel{3}{$\ket{\varphi(0.x_0x_1x_2x_3)}$}
  \end{qcircuit}
  \hspace{40pt} $\overrightarrow{\rule{30pt}{0pt}}$
  \begin{qcircuit}{24}{0}{0}{0}
    \qmultiple{0}{1}{$\widetilde{\mathcal{F}}$}
  \end{qcircuit}
\end{center}

Following an idea of Draper \cite{qadd}, the accumulation of
information into the phase coefficients can continue using conditional
phase shifts from a second register into the Fourier transformed one.
The state of the first register after this stage is the
$\widetilde{\mathcal{F}}$ transformed state of \ket{x+y} (modulus $2^4$).
It is easy to see that all the phase shifts of the same kind involve
independent qubit lines, and can therefore be represented by a single
time slice.

\begin{center}
  \begin{qcircuit}{12}{5}{8}{70}
    \qlabel{4}{\ket{y_3}} \qlabel{5}{\ket{y_2}}
    \qlabel{6}{\ket{y_1}} \qlabel{7}{\ket{y_0}}
    \qlabel{0}{$\ket{\varphi(0.x_3)}$}
    \qlabel{1}{$\ket{\varphi(0.x_2x_3)}$}
    \qlabel{2}{$\ket{\varphi(0.x_1x_2x_3)}$}
    \qlabel{3}{$\ket{\varphi(0.x_0x_1x_2x_3)}$}
    \qcondrot{3}{7}{1} \qsame[0.8] \qcondrot{2}{6}{1} \qsame[0.8] 
    \qcondrot{1}{5}{1} \qsame[0.8] \qcondrot{0}{4}{1}
    \qcondrot{3}{6}{2} \qsame[0.8] \qcondrot{2}{5}{2} \qsame[0.8]
    \qcondrot{1}{4}{2}
    \qcondrot{3}{5}{3} \qsame[0.8] \qcondrot{2}{4}{3}
    \qcondrot{3}{4}{4}
    \qendlabel{4}{\ket{y_3}} \qendlabel{5}{\ket{y_2}}
    \qendlabel{6}{\ket{y_1}} \qendlabel{7}{\ket{y_0}}
    \qendlabel{0}{$\ket{\varphi(0.x_3 + 0.y_3)}$}
    \qendlabel{1}{$\ket{\varphi(0.x_2x_3 + 0.y_2y_3)}$}
    \qendlabel{2}{$\ket{\varphi(0.x_1x_2x_3 + 0.y_1y_2y_3)}$}
    \qendlabel{3}{$\ket{\varphi(0.x_0x_1x_2x_3 + 0.y_0y_1y_2y_3)}$}
  \end{qcircuit}
  \hspace{0pt} $\overrightarrow{\rule{30pt}{0pt}}$
  \begin{qcircuit}{24}{0}{0}{0}
    \qmultiple{0}{3}{$\mathcal{R}$}
  \end{qcircuit}
\end{center}

The obvious step now is to apply $\widetilde{\mathcal{F}}^\dagger$ to
$\widetilde{\mathcal{F}}\ket{x+y}$ in order to get the modular addition
of $x$ and $y$ (see the left half in the following picture). Once the
adder circuit $\mathcal{A}$ is constructed, the process can be iterated
in order to build a three-input adder, by summing the content of a third
register onto the register which holds the intermediate sum. The right
half in the following figure shows the resulting circuit, where
$\mathcal{A}_1$ and $\mathcal{A}_2$ represent the $\mathcal{A}$ operator
acting on the first and third registers.

\begin{center}
  \begin{qcircuit}{20}{3}{2}{20}
    \qlabel{0}{\ket{y}} \qlabel{1}{\ket{x}}
    \qmultiline{0}{5} \qmultiline{1}{5}
    \qsingle{0}{$\widetilde{\mathcal{F}}$}
    \qmultiple{0}{1}{$\mathcal{R}$}
    \qsingle{0}{$\widetilde{\mathcal{F}}^{\;\mydagger}$}
    \qendlabel{0}{\ket{x\!+\!y}}\qendlabel{1}{\ket{x}}
  \end{qcircuit}
  \hspace{-10pt}
  $\overrightarrow{\rule{0pt}{5pt}\rule{30pt}{0pt}}$
  \begin{qcircuit}{20}{0}{0}{0}
    \qmultiple{0}{1}{$\mathcal{A}$}
  \end{qcircuit}
  \hfill
  \begin{qcircuit}{20}{2}{3}{20}
    \qlabel{0}{\ket{z}} \qlabel{1}{\ket{y}} \qlabel{2}{\ket{x}}
    \qmultiline{0}{5} \qmultiline{1}{5} \qmultiline{2}{5}
    \qmultiple{0}{1}{$\mathcal{A}$} \qcontrol{0}{2}{$\mathcal{A}_2$}
    \qsame \qsingle{2}{$\mathcal{A}_1$}
    \qendlabel{0}{\ket{x\!+\!y\!+\!z}}
    \qendlabel{1}{\ket{y}} \qendlabel{2}{\ket{x}}
  \end{qcircuit}

\end{center}

The operator syntax of the proposed quantum language allows to write
source code for the implementation of the three-input adder which
strictly follows the previous ``high level'' description. The
desired circuit is built by the following function with \texttt{size}=4.

\begin{verbatim}
Qop build_three_adder(int size) {
   Qop phase_shifts;
   for (int i=0; i<size; ++i)
     phase_shifts << QCondPhase(size-i, i+1).offset(i);
   Qop transform = (QFourier(size) & QSwap(size)).offset(size);
   Qop adder_2   = transform & phase_shifts & (! transform);
   Qop adder_3   = (adder_2 >> size);
   adder_3 << adder_2.split(size, size);
   return adder_3;
}
\end{verbatim}
The loop sets up the $R$ circuit into the \texttt{phase\_shifts}
operator, which is initialised to the identity, by pushing the
conditional phase shifts into it. The first argument to each
\texttt{QCondPhase} is the number of gates to be stored and the
second is the power $k$ of $C_{R_k}$ (see \ref{sec:lowlevel}).
Each \texttt{QCondPhase} is then offset to the correct position.
The \texttt{transform} operator contains the Fourier transform over
the lower register (\texttt{offset(size)}) once the final qubit swap
has been reversed with a \texttt{QSwap} operator. \par
\texttt{phase\_shifts} is then combined with \texttt{transform} and
its adjoint to form the two-input adder \texttt{adder\_2}. The
three-input adder \texttt{adder\_3} is built by concatenating two
permutations of \texttt{adder\_2}; the first is offset by
\texttt{size} qubit lines (thus it acts on the second and third
register) and the second is split with a hole in the middle
(thus it acts on the first and third registers). Note that the
\texttt{<<} operator at the end reuses the time slices in
\texttt{adder\_2}, which is therefore lost. \par
During the construction of \texttt{transform} any trivial
simplification algorithm embedded in the language 
(see section \ref{sec:opcomp}) can simplify the double swapping of
qubit lines at the end of the Fourier transform. The same algorithm
can simplify the $\widetilde{\mathcal{F}}^\dagger$ at the end of the
first two-input adder $\mathcal{A}$ with the $\widetilde{\mathcal{F}}$
at the beginning of the second adder (since they are performed on the
same register). The data which is stored inside the quantum operator 
object \texttt{adder\_3} is therefore the following, where quantum
gates which belong to the same time slice have been grouped together,
reducing the number of time slices to $28$:
\medskip

\begin{qcircuit}{10}{30}{12}{0} \scriptsize
  \qlabel{0}{\hspace{-20pt}(11)} \qlabel{1}{\hspace{-20pt}(10)}
  \qlabel{2}{\hspace{-20pt}(9)}  \qlabel{3}{\hspace{-20pt}(8)}
  \qlabel{4}{\hspace{-20pt}(7)}  \qlabel{5}{\hspace{-20pt}(6)}
  \qlabel{6}{\hspace{-20pt}(5)}  \qlabel{7}{\hspace{-20pt}(4)}
  \qlabel{8}{\hspace{-20pt}(3)}  \qlabel{9}{\hspace{-20pt}(2)}
  \qlabel{10}{\hspace{-20pt}(1)} \qlabel{11}{\hspace{-20pt}(0)}
  \qevidence[\texttt{transform}]{-1}{10}
  \qhadamard{3} \qcondrot{3}{2}{2} \qcondrot{3}{1}{3} \qcondrot{3}{0}{4}
  \qhadamard{2} \qcondrot{2}{1}{2} \qcondrot{2}{0}{3}
  \qhadamard{1} \qcondrot{1}{0}{2}
  \qhadamard{0}
  \qskip[0.2]
  \qevidence[\texttt{phase\_shifts}]{-1}{5}
  \qcondrot{3}{7}{1} \qsame[0.8] \qcondrot{2}{6}{1} \qsame[0.8]
  \qcondrot{1}{5}{1} \qsame[0.8] \qcondrot{0}{4}{1}
  \qcondrot{3}{6}{2} \qsame[0.8] \qcondrot{2}{5}{2} \qsame[0.8]
  \qcondrot{1}{4}{2}
  \qcondrot{3}{5}{3} \qsame[0.8] \qcondrot{2}{4}{3}
  \qcondrot{3}{4}{4}
  \qevidence[\texttt{phase\_shifts}]{-1}{5}
  \qcondrot{3}{11}{1} \qsame[0.8] \qcondrot{2}{10}{1} \qsame[0.8]
  \qcondrot{1}{9}{1} \qsame[0.8] \qcondrot{0}{8}{1}
  \qcondrot{3}{10}{2} \qsame[0.8] \qcondrot{2}{9}{2} \qsame[0.8]
  \qcondrot{1}{8}{2}
  \qcondrot{3}{9}{3} \qsame[0.8] \qcondrot{2}{8}{3}
  \qcondrot{3}{8}{4}
  \qevidence[\texttt{transform} (adjoint)]{-1}{10}
  \qhadamard{0}
  \qcondrot{1}{0}{2^{\mydagger}} \qhadamard{1}
  \qcondrot{2}{0}{3^{\mydagger}}
  \qcondrot{2}{1}{2^{\mydagger}} \qhadamard{2}
  \qcondrot{3}{0}{4^{\mydagger}}
  \qcondrot{3}{1}{3^{\mydagger}}
  \qcondrot{3}{2}{2^{\mydagger}} \qhadamard{3}
\end{qcircuit}

\subsection{An example with phase estimation}

This example is an implementation of the phase estimation algorithm
as a subroutine of the randomised order finding algorithm, in order
to illustrate the use of constructors for controlled operators.
The order finding algorithm computes the order%
\footnote{For positive integers $x$ and $N$, with no common factors,
  the order of $x$ modulo $N$ is defined to be the least positive
  integer $r$ such that $x^r \textrm{~mod~} N = 1$.}
$r$ of $x$ with respect to $N$, where $x$ and $N$ are two coprime
integer variables with $x<N$. The phase estimation subroutine is used
to return a mantissa which approximates $s/r$ where $s$ is a random
number in $[0,\ldots,r-1]$. The result is to be passed through the
continued fraction algorithm (which is completely classical) in order
to extract $r$. The interested reader can find further details in
\cite{QAR}. \par
The phase estimation subroutine accepts two additional parameters,
$\epsilon$ and $n$, where $1-\epsilon$ is the probability bound on
having $n$ exact digits in the decimal expansion of $s/r$.
The corresponding circuit is the following, where $M(q)$ is a matrix%
\footnote{The definition of $M(q)$ is the following: $M(q) \ket{i}
  = \ket{iq \textrm{~mod~} N}$ if $i<N$, the identity otherwise.
  If $q$ and $N$ are coprime the corresponding transformation is indeed
  invertible hence $M(q)$ is unitary. There are various strategies for
  the implementation of $M(q)$; the simplest one uses classical
  preprocessing, controlled summations and ancilla qubits. It would
  make the example too complicated to show the actual construction of 
  $M(q)$, which is however detailed by many authors. See for instance
  Shor \cite{shor}.}
which implements the multiplication by $q$ modulo $N$, and $M_j$
stands for $M(x^{2^j})$
\begin{center}
  \begin{qcircuit}{15}{5}{6}{70}
    \qlabel{0}{{\small eigen. register}}
    \qlabel{3}{{\small phase register}}
    \qbrace{0}{2}{\ket{1}}{$m$ qubits}
    \qbrace{3}{5}{\ket{0}}{$t$ qubits}
    \qpatchline{1}{0}{6}{3 5} \qpatchline{4}{0}{6}{3 5}
    \qmultiple{3}{5}{H} \qmulticontrol{0}{2}{5}{}
    \qsame \qevidence[$M_0$]{1}{1} \qskip
    \qskipping{0}{4} \qmulticontrol{0}{2}{3}{}
    \qsame \qevidence[$M_{t\textrm{-}\!1}$]{1}{1} \qskip
    \qmeasure{3} \qsame \qmeasure{4} \qsame \qmeasure{5}
    \qendbrace{3}{5}{estimate of $\frac{s}{r}$}
    \rmove(70 0)
  \end{qcircuit}
\end{center}
\begin{verbatim}
Qbitset run_order_finding(int x, int N, int n, float epsilon) {
   int t = n + ceil(log(1+1/(2*epsilon))/log(2));
   int m = ceil(log(N)/log(2));
   int q = x;
   Qop controlled_multiply[t];
   for (int i=0; i<t; ++i, q = ((q*q) % N))
      controlled_multiply[i] << Qop(generate_multiply(q, N), 1);
   Qop mixer = QHadamard(t);
   Qreg phase(t);
   Qreg eigen(m, 1);
   mixer(phase);
   for (int i=0; i<t; i++)
      controlled_multiply[i](phase[i] & eigen);
   return phase.measure();
}
\end{verbatim}
The first lines simply calculate the number \texttt{t} of qubits
in the phase register and the size \texttt{m} of the eigenvector
register needed in order to host \texttt{N}.
Then, for each power $q\in\{x^1,\cdots,x^{2^t}\}$, the helper function
\texttt{generate\_multiply} builds the $M(q)$ operator. The returned
object (which is a \texttt{Qop}) is immediately used as first argument
of the controlled operator constructor in order to build a one-qubit
controlled $M(q)$ for later use (this last operator acts on
\texttt{m}+1 sized registers). \par
Everything up to now is classical preprocessing, the interaction
with the quantum device starts with the creation and initialisation
of the phase and eigenvector registers, followed by the application
of the tensor product of \texttt{t} Hadamard gates (\texttt{mixer})
to the phase register, transforming its state into the uniform
superposition of all computational basis states. \par
Inside the main loop the controlled multiplications are then executed
by passing the control qubit and the target register together (using
the register concatenation operator). The last line measures the
phase register and returns the phase estimate as a bit set.

\subsection{An example with Grover's algorithm}
\label{sec:grover}

This example is the well known Grover's algorithm \cite{grover}
which finds one or more elements in an unstructured input space with
an exponential speed-up with respect to the classical case. It is
meant to illustrate the usefulness of having a HLP for the automatic
construction of oracle operators from functions specified with the
underlying classical language. The premise is that an efficient
algorithm is known for the computation of the classical oracle
function $f$. The following example is the simplified version in
which there is exactly one good input marked by the oracle, that is
$f(x)=$ \texttt{true} if and only if $x$ is the searched element
$\tilde{x}$. The range of inputs is $[0,\ldots,N-1]$ where $N=2^n$.
The corresponding circuit is the following, where $O$ is the phase
oracle operator, $M$ is the so called ``inversion about mean'' and
$G=OM$ is the Grover iteration:
\medskip
\begin{center}
  \begin{qcircuit}{20}{5}{1}{20}
    \qmultiline{0}{5} \qlabel{0}{\ket{0}}
    \qhadamard{0} \qevidence[$\sqrt{N}$ times]{-1}{3}
    \qskip \qsingle{0}{$G$} \qskip \qmeasure{0}
    \qendbrace{0}{0}{$\tilde{x}$ with high probability}
    \rmove(80 0)
  \end{qcircuit}
\end{center}
\begin{verbatim}
Qbitset run_Grover(bool(*f)(int), int n) {
   int repetitions = sqrt(pow(2.0,n));
   Qop phase_oracle(f,n);
   Qop invert_zero(f_0,n);
   Qop mixer = QHadamard(n);
   Qop invert_mean = mixer & invert_zero & mixer;
   Qop grover_step = phase_oracle & invert_mean;
   Qreg input(n);
   mixer(input);
   for (int i=0; i<repetitions; ++i) grover_step(input);
   return input.measure();
}
\end{verbatim}
At the beginning the number of iterations to be performed is
calculated. It is well known that this number scales as
$O(\sqrt{N})$ \cite{grover}. Then the \texttt{phase\_oracle} and
\texttt{invert\_zero} operators are built (\texttt{f\_0} is a
function which returns \texttt{true} only when the input is $0$).
This construction relies on automatic translation from the
corresponding classical function provided by the language. This is
a major difficulty in the language implementation and is discussed
in section \ref{sec:pseudoclassical}. \par
Once the previous two operators are ready, it is a matter of
composition to build up the \texttt{invert\_mean} (inversion about
mean) and the \texttt{Grover\_step}. Note that up to now everything
is classical preprocessing. The quantum part of the routine starts
when an \texttt{input} register is created with the appropriate
size for holding the input range; then this register is subject
to a Hadamard gate on each qubit line in order to generate the
uniform superposition of all possible inputs. 
When the \texttt{input} is ready, the \texttt{Grover\_step} is
applied \texttt{repetitions} times, the iteration counter being
classical. The algorithm is terminated by a register measurement
which returns $\tilde{x}$ with high probability.

\section{Language internals}
\label{sec:internals}

\subsection{Operator composition and simplification}
\label{sec:opcomp}

This section analyses with more details the benefits gained by
using quantum operator objects instead of functions for representing
quantum circuits; the function-like approach, basically, is adopted
both in the \qcl \cite{QCL1, QCL2} and in the \qgcl
\cite{zuliani_programming} languages. \par\noindent
\picinpar
{\quad\quad A simple example can stress the difference between the
  two solutions: a \texttt{Hadamard\_2()} function is available, which
  accepts a quantum register and an index $i$ inside the register as
  arguments. It applies two Hadamard gates, to the $i$-th and to the
  $i+1$-th elements of the register, and is invoked with $i$ assuming
  all the possible values for a valid index in the register:}
{\begin{qcircuit}{10}{5}{6}{0} \tiny
    \qlabel{6}{} 
    \qhadamard{5} \qsame \qhadamard{4}
    \qhadamard{4} \qsame \qhadamard{3}
    \qhadamard{3} \qsame \qhadamard{2}
    \qhadamard{2} \qsame \qhadamard{1}
    \qhadamard{1} \qsame \qhadamard{0}
  \end{qcircuit}%
  $\Longrightarrow\!\!\!$%
  \begin{qcircuit}{10}{1}{6}{0} \tiny
    \qlabel{6}{} \qhadamard{5} \qsame \qhadamard{0}
  \end{qcircuit}}
\begin{quote}
\begin{verbatim}
  Qreg myreg(size);
  for (int i=0; i<(size-1); i++) Hadamard_2(myreg, i);
\end{verbatim}
\end{quote}
The insert on the right shows the corresponding circuit (for
\texttt{size}$=6$) and its obvious simplification. It is clear that
the quantum language should perform this optimisation, transparently
to the user. With a function-like syntax however the quantum code
is generated independently by each function call and the optimisation
can be done only if the code is buffered and simplified before being
sent to the quantum device. Moreover, if the whole loop is repeated
with a different register, the buffering and the simplification have
to be redone, even though the optimisation depends only on the circuit
structure and not on the actual register. \par
What is really needed is a mechanism for generating the whole
circuital description before the quantum device is even fired up,
applying algebraic simplifications {\em once for all}. This is possible
if quantum operators are implemented as data structures modifiable
at run time, which can be manipulated, composed and simplified
{\em before} allocating the quantum registers. The simplification
routines which perform optimisations can be embedded%
\footnote{If some optimisation routines are too expensive for being
  embedded, it is possible to leave to the programmer the freedom to
  force their call, e.g. \texttt{circuit.simplify(\dots)} where the
  arguments select the simplification strategy.}
inside the composition primitives. In the proposed language the
previous example is coded as:
\begin{quote}
\begin{verbatim}
  Qop circuit;
  for (int i=0; i<(size-1); i++) circuit << QHadamard(2).offset(i);
  Qreg myreg(size);
  circuit(myreg);
\end{verbatim}
\end{quote}

\subsection{The implementation of classical functions}
\label{sec:pseudoclassical}

A requirement for a useful quantum language is the ability to
implement pseudo-classical operators, that is transformations
like $U_f : \ket{x}\ket{y} \rightarrow \ket{x}\ket{y \oplus f(x)}$
where $f: \mathbb{Z}_{2^n} \rightarrow \mathbb{Z}_{2^m}$ is a
classical function and $n$ and $m$ are the sizes of the registers.
In section \ref{sec:qops} it was suggested the introduction of an
HLP for the automatic construction of quantum operators from the
classical specification of an algorithm for $f$; this section
discusses the problem more extensively. \par
Lecerf \cite{lecerf} and Bennett \cite{bennett} have shown
that any classical, potentially irreversible, algorithm can be
efficiently converted into a reversible one. Since reversible
classical algorithms can be converted into equivalent quantum
operators efficiently, this means that each classical algorithm
computing a function $f$ can be efficiently converted into a
quantum operator $U_f$. A constructive approach to this problem
has been shown by Zuliani \cite{zuliani_pseudoclassical}. \par
What is to be remarked immediately is that if $f$ is ``known''
through a classical black-box it is useless as far as quantum
computing is concerned. The reason for this is that getting the
action of $f$ through a black-box requires as many queries as the
size of the input space; hence, if the classical preprocessing
stage tries to understand $f$ through a black-box, it experiences
an exponential slowdown which nullifies any quantum gain. 
Therefore the constructor of a pseudo-classical operator does not
need the ability to {\em call} the function but the ability to
{\em inspect} its algorithmic definition. \par
A second remark is that ``function'' in the current context means
a mathematical function, i.e. a deterministic mapping of any input
to an output. This definition is more restrictive than the usual
meaning in a programming language (a routine), because a routine
may depend on the state of the classical machine%
\footnote{The dependence on the state of the classical machine
  can be through global or static variables, including random
  number generators, as well as through run time conditions,
  like user input}
and may be not terminating on some inputs. \par
Summarising, the automatic generation of pseudo-classical quantum
operators needs access to the specification of the algorithm which
implements the function. Moreover, either the algorithm is written
in a restricted language which allows only the coding of mathematical
mappings, or a ``filter'' must be applied in order to check for the
presence of operations which depend on the state of the classical
machine or which may cause the algorithm not to terminate.
The filtered algorithm must then be parsed and transformed into a
finite size circuit. No scheme for this translation has yet been
developed for the proposed quantum language.

\section*{Conclusions}

It has been proposed a language scheme and a set of high level
primitives for programming a \qram machine. The high level
primitives have been studied in order to fit with current circuit
model descriptions of quantum algorithms. \par
The scheme provides an automatic translation and optimisation of
high level primitives into low level primitives which are sent to
the quantum device. This generated code is still hardware
independent, in order to make it easy to switch from real quantum
devices to quantum simulators and between different models of
quantum hardware and different schemes for low level hardware
dependent primitive translation. \par
There is an ongoing effort%
\footnote{A detailed description of a preliminary language
  implementation can be found in: S.Bettelli, PhD thesis,
  University of Trento, in preparation (February 2002).
  The code will be made available at: \\
  \texttt{http://sra.itc.it/people/serafini/quantum-computing/qlang.html}}
to provide a working implementation of the ideas indicated in
this paper through a library using the \cpp programming language. 
A procedure for the automatic translation of classical mappings
(see \ref{sec:pseudoclassical}) is still to be studied. \par
Once this task is accomplished, it could be a valuable tool for
a number of different purposes, like:
\begin{itemize}
\item testing the efficiency of different high level simplification
  and optimisation routines for quantum circuits, including the
  implementation of pseudo-classical operators;
\item testing the efficiency of different schemes for high level
  to low level and hardware independent to dependent translation 
  routines for quantum circuits;
\item testing the efficiency of different hardware architectures
  for the execution of quantum code (with timing simulations);
\item having an high level interface for the specification of
  algorithms which are to be fed into quantum simulators;
\item testing the robustness of error correction codes and fault
  tolerant quantum computation with respect to generic error
  models, without modifying the simulation libraries;
\item quantum programming (when quantum computers will be ready)
\end{itemize}
\bigskip
  The authors wish to thank Bruno Caprile (ITC-IRST) for interesting
  discussions about the design and the aims of the programming language.
  S.B. was a doctoral student at the University of Trento, also 
  associated to INFN, during the preparation of this work.

\appendix

\section{Implementation details}

\subsection{A detailed scheme for the language implementation}
\label{app:implementation}

\begin{figure}[!t]
  \begin{center}
    \includegraphics[width=.8\linewidth]{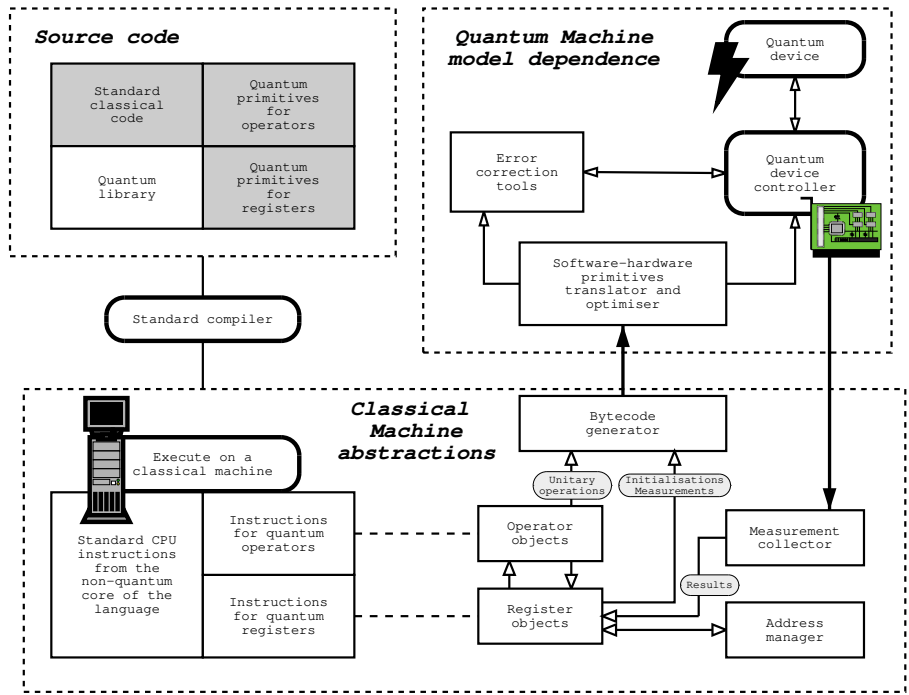}
    \caption{Overall scheme for a quantum device controlled by
      classical hardware, which details fig.\ref{fig:overview}.
      The three shaded boxes are the resources available to the
      programmer for writing a ``quantum program''. The big dashed
      boxes contain the source code level, the classical machine
      control and the quantum machine architecture dependence.
      See appendix \ref{app:implementation} for more details.}
    \label{fig:scheme}
  \end{center}
\end{figure}

In section \ref{sec:implementation}, the quantum registers and the
quantum operators were introduced. Their syntax was examined in
section \ref{sec:regs} and \ref{sec:qops}. This appendix takes a
closer look at these data types and at the language environment
by describing the scheme presented in fig.\ref{fig:scheme}. \par
The specification of a ``quantum program'' starts with a source
code text file, just like a plain program. The source code syntax
uses a standard programming language as a base, and adds primitives
for creating and managing quantum {\em register} objects and quantum 
{\em operator} objects. Additional routines (a ``quantum library'')
for common circuits may be used. The code is compiled to an
executable by a standard compiler for the base language. \par
At run-time this executable creates in the classical memory some
data structures which correspond to the operator and register
objects, and manages their ``interaction''. The data structures
for quantum registers are basically lists of distinct addresses. 
The implementation of non-unitary operations (initialisations
and measurements) is achieved directly through these
interfaces. \par
The ``usage count'' of each qubit (the number of registers which
are referencing it) is kept by another data structure, the
{\em address manager}, which can not be directly manipulated
by the programmer. The address manager knows which qubits are
``free'' and provides lists of free addresses with the appropriate
size when a new register is to be created. An example of a set of
overlapping quantum registers with the corresponding status of the
address manager was shown in fig.\ref{fig:registers}. \par
Quantum operator objects, when applied onto registers, calculate
which gates are to be executed on which qubit addresses and send
this information to the {\em byte-code generator}, which provides
an additional address translation in order to perform qubit swaps
without resorting to the quantum device; an approach for these
calculations is shown in appendix \ref{app:translation}. \par
The byte-code generator interfaces directly to a specific (hardware
dependent) quantum device driver, exporting a stream of quantum
gate codes and the locations where they must be executed. Quantum
gate codes are still hardware independent: the translation to
the real hardware primitives takes place at this stage. This
allows for a very simple way to substitute an emulator to the
real device. The device driver can implement additional specific
optimisations and error correction tools. \par
What is appealing here is that all this machinery can be
implemented by using a set of libraries and a standard compiler
for an object oriented language. Our group has produced a
prototype for these ideas using the \cpp \cite{C++} language.

\subsection{Implementation of controlled circuits}
\label{app:controlled}

This appendix introduces a possible approach for the construction
of multi-controlled circuits. Though it is not part of the language
definition and, to some extent, dependent on a particular choice for
the elementary gate set, this approach shows that multi-controlled
operators can be implemented with the same space and time complexity
as the corresponding uncontrolled ones. Many ideas in the following
are taken from the classic paper by Barenco et al. \cite{gates} and
extended with the notion of parallelisation of homogeneous gates
introduced in section \ref{sec:lowlevel}. \par
First, one needs to recognise that the controlled version of each
gate in the chosen set of elementary gates $\{ H, R_k, C_{R_k} \}
_{k\in \mathbb{N}}$ can be implemented by a circuit with depth bounded
by a global constant. The controlled $R_k$ is $C_{R_k}$ itself, which
is a primitive, so that only the construction of $C_H$ and $C_{C_{R_k}}$
has to be shown. \par
The construction of $C_H$ is quite easy; first, $H$ is decomposed
into a sequence of three rotations around the $z$, $x$ and $z$ axis 
(Euler angles decomposition, though usually the chosen axes are $z$
and $y$):
\begin{displaymath}
  H = i \, R_z\!\left(\frac{\pi}{2}\right)
  R_x\!\left(\frac{\pi}{2}\right) R_z\!\left(\frac{\pi}{2}\right)
\end{displaymath}
Since $HZH = X$, the $R_x$ rotation can be turned into a $R_z$ 
rotation with the same argument between two Hadamard matrices. 
The $R_z(\frac{\pi}{2})$ matrix is the same as 
$e^{-i\frac{\pi}{4}}R_2$, hence:
\begin{displaymath}
  H = i \left( e^{-i\frac{\pi}{4}} \right)^3
  R_2 H R_2 H R_2 = e^{-i\frac{\pi}{4}} R_2 H R_2 H R_2
\end{displaymath}
The previous relation can be turned into a circuit for $C_H$ by
controlling the three phase shifts (since $H^2$ is the identity)
and providing the phase factor with a phase shift $R^\dagger_3$
on the control line:
\begin{center}
  \begin{qcircuit}{15}{1}{2}{0}
    \qcontrol{0}{1}{H}
  \end{qcircuit}
  $\longmapsto$
  \begin{qcircuit}{15}{6}{2}{0}
    \qcondrot{0}{1}{2} \qhadamard{0} \qcondrot{0}{1}{2} \qhadamard{0}
    \qcondrot{0}{1}{2} \qrot{1}{3^{\mydagger}} 
  \end{qcircuit}
\end{center}
The doubly controlled phase shift $C_{C_{R_k}}$ can be built by
adapting a circuit known in literature%
\footnote{See fig.4.9 on page 182 in \cite{textbook}}
for the Toffoli gate (the symbols $k_1$ and $k_2$ stand for $k+1$
and $k+2$ respectively):

\begin{center}
  \begin{qcircuit}{15}{1}{3}{0}
    \qcondrot{0}{1}{k} \qsame \qsegment{1}{2}
  \end{qcircuit}
  $\longmapsto$
  \begin{qcircuit}{15}{17}{3}{0}
    \qhadamard{0} \qcondrot{0}{1}{1} \qhadamard{0} \qrot{0}{k_2^{\mydagger}}
    \qhadamard{0} \qcondrot{0}{2}{1} \qhadamard{0} \qrot{0}{k_2}
    \qhadamard{0} \qcondrot{0}{1}{1} \qhadamard{0} \qrot{0}{k_2^{\mydagger}}
    \qhadamard{0} \qcondrot{0}{2}{1} \qhadamard{0} \qrot{0}{k_2}
    \qcondrot{1}{2}{k_1}
  \end{qcircuit}
\end{center}

This construction could be generalised to multi-controlled phase shifts,
but the depth would scale exponentially with the number of controls.
It is easy to see that this circuit performs $C_{C_{R_k}}$ correctly.
Whenever one of the control qubits is found in the \ket{0} state, all
the gates on the target line cancel out and $C_{R_{k + 1}}$ between the
controls has no effect. When the control qubits are found in \ket{11}
the following relation holds:

\begin{displaymath}
  X R_{k+2} = \left(\begin{array}{cc}0&\phi_{k+2}\\1&0\end{array}\right)
  \quad \Longrightarrow \quad
  A = X R_{k+2}^\dagger X R_{k+2} = \phi_{k+2}^* R_{k+1}
  \quad \Longrightarrow \quad
  \phi_{k+1} A^2 = R_k
\end{displaymath}

The depth of this circuit is a constant, independently from the
parameter $k$. A Toffoli gate \label{ctor:toffoli} (doubly controlled
NOT) can be obtained by enclosing a $C_{C_{R_1}}$ with two Hadamard
matrices on the target line, since $HR_1H=HZH=X$. For the same reason,
the CNOT gate \label{ctor:cnot} can be built using $C_{R_1}$. \par
\begin{figure}
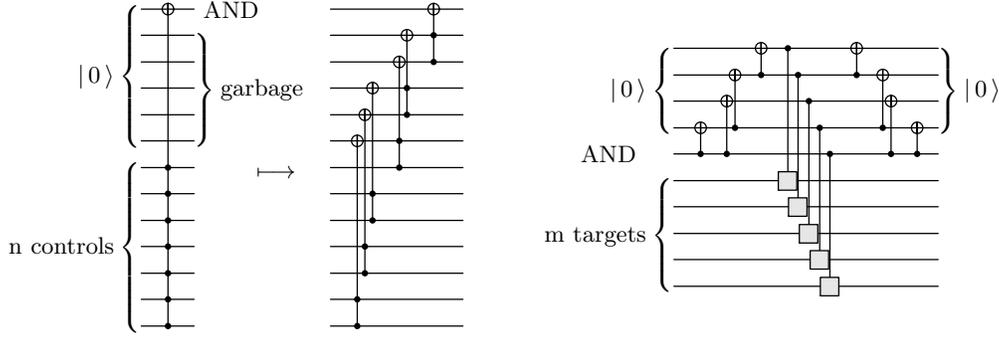

  \begin{center} \small
    ~\hfill
    \begin{qcircuit}{10}{1}{13}{35} 
      \qbrace{0}{6}{}{n controls} \qbrace{7}{12}{}{\ket{0}}
      \qendlabel{12}{AND} \qendbrace{7}{11}{garbage}
      \qtoffoli{12}{6}{5} \qsame \qsegment{5}{4} \qsegment{4}{3}
      \qsegment{3}{2} \qsegment{2}{1} \qsegment{1}{0}
    \end{qcircuit}
    \hspace{-30pt} $\longmapsto$
    \begin{qcircuit}{10}{4}{13}{0}
      \qtoffoli{7}{1}{0} \qsame[0.7] \qtoffoli{8}{3}{2} \qsame[0.7]
      \qtoffoli{9}{5}{4} \qtoffoli{10}{7}{6} \qsame[0.7]
      \qtoffoli{11}{9}{8} \qtoffoli{12}{11}{10}
    \end{qcircuit}
    \hfill
    \begin{qcircuit}{10}{9}{10}{35}
      \qlabel{5}{AND} \qbrace{6}{9}{}{\ket{0}}
      \qbrace{0}{4}{}{m targets}
      \qnot{6}{5} \qnot{7}{5} \qsame[0.7] \qnot{8}{6} \qnot{9}{8}
      \qcontrol{4}{9}{} \qsame[0.6] \qcontrol{3}{8}{} \qsame[0.6]
      \qcontrol{2}{7}{} \qsame[0.6] \qcontrol{1}{6}{} \qsame[0.6]
      \qcontrol{0}{5}{}
      \qnot{9}{8} \qnot{8}{6} \qsame[0.7] \qnot{7}{5} \qnot{6}{5}
      \qendbrace{6}{9}{\ket{0}}
    \end{qcircuit}
    \hfill~
  \end{center}
  \caption[Implementation of controlled operations]{
    (a) The circuit on the left shows the implementation of the
    coincidence circuit, that which calculates the AND of all
    the $n$ control lines. $n-1$ ancilla qubits prepared in the
    \ket{0} state are needed, including that which holds the
    result of the computation. If independent Toffoli gates are
    performed in parallel, the circuit depth grows like $\log n$.
    The adjoint of the circuit must be applied at the end of the
    controlled circuit for properly uncomputing the ancillae.
    (b) The circuit on the right shows how the qubit holding the
    AND of all the controls can be ``copied'' enough times to
    allow for a parallel implementation of $m$ independent
    operations; indeed this is not a copy but the transformation
    $(\alpha\ket{0} + \beta\ket{1})\otimes\ket{0\dots0}
    \rightarrow \alpha\ket{0\dots0} + \beta\ket{1\dots1}$
    which leads to a many-particle entangled state.
    The circuit of course requires $m-1$ additional qubits, to be
    uncomputed at the end. The depth of the copying and uncopying
    section grows like $\log m$.
  }
  \label{fig:controlled}
\end{figure}

With $n-1$ Toffoli gates and $n-1$ ancilla qubits prepared in the
\ket{0} state it is possible to calculate the AND of a $n$-qubit
register; if independent Toffoli gates can be applied in parallel
the circuit depth grows like $\log n$. The construction is optimal
when $n$ is a power of two%
\footnote{This is because the first bunch of Toffoli gates calculates
  $n/2$ ANDs, the second bunch half of that and so on, until the
  number of gates is one, hence $n/2^q \sim 1$ where $q$ is the number
  of steps.}.
An example of this coincidence circuit for $n=7$ is shown in figure
\ref{fig:controlled}. \par

Once the qubit holding the (quantum) AND of all the controls is ready,
it can be used to perform the controlled operations. Since however
it is unlikely that a single physical system could be used to control
at the same time a number of different qubit lines, this would prevent
parallelisation in the controlled operator, changing its complexity.
A workaround consists in ``copying'' the single control into $m$
qubits, $m$ being the maximum number of parallel gates in a single
time slice of the uncontrolled operation. Indeed this consists in the
transformation $(\alpha\ket{0} + \beta\ket{1})\otimes\ket{0\dots0}
\rightarrow \alpha\ket{0\dots0} + \beta\ket{1\dots1}$, which has a
logarithmic complexity%
\footnote{During the first step the control is used to perform one
  copy, then the two qubits can be used during the second step for
  two copies and so on, hence $\sum_{j=0}^{q-1} 2^j \sim 2^q \sim m$
  where $q$ is the number of steps and scales as $\log m$.}.
Each of the $m$ independent controlled operations can then be
performed using a different qubit as control (see figure
\ref{fig:controlled}). \par
The size of the register to be fed into the controlled operator is
$m+n$ ($n$ controls and $m$ targets). The number of additional
qubits to be used as ancillae is $m+n-2$, therefore the ratio between
the space requirements and the operator ``size'' is less than 2.
The complexity of the calculation and uncomputation of the AND of
all the controls is $\log n$; that of the control copy is $\log m$.
These values have to be compared with a (likely) polynomial complexity
in $m$ for the uncontrolled operator.

\subsection{Techniques for managing qubit addresses}
\label{app:translation}

\begin{figure}
  \begin{center}
    \includegraphics[width=.8\linewidth]{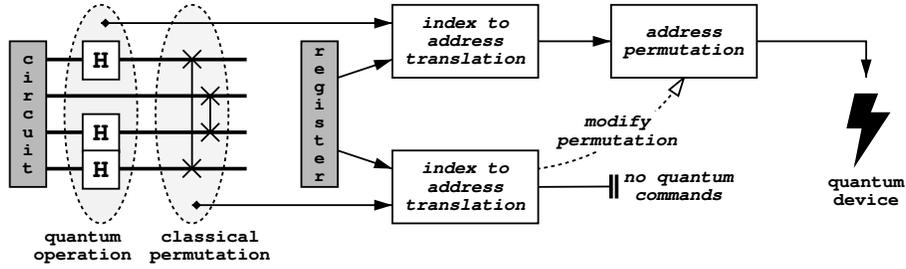}
    \caption[Implementation of register permutations]{
      This figure visualises the difference between the treatment
      of a real quantum operation and a classical permutation. In
      the first case (a time slice of Hadamard matrices) the index
      list in the time slice is used to select some of the addresses
      contained in the quantum register object. These addresses are
      then subject to a further translation (a permutation) before
      being fed into the quantum device. In the second case (a swap
      time slice) the run time environment runs only the first
      translation, then uses the result in order to modify the
      permutation function for the following time slices. This
      change affects all the registers and is thus indistinguishable
      from a hardware swap for what concerns the programmer.}
    \label{fig:translation}
  \end{center}
\end{figure}

These appendix, and the scheme in figure \ref{fig:translation}, detail
a possible approach for managing qubit addresses. The first part
explains how the specifications of circuits and registers are matched
to calculate which qubit locations the LLP must be executed on. The
second part shows how to implement qubit line swaps as classical
operations instead of as hardware ones, as suggested on page
\pageref{sec:qops:reorder}. \par
Quantum operators (sec.\ref{sec:qops} and \ref{sec:lowlevel}) are
stored as sequences of time slices, each of which is specified by one
or more index lists. Each quantum register (sec.\ref{sec:regs}) is
specified by a list of addresses. Therefore, there is a common basic
data structure, a ``list'', which can be implemented by ordered sets
of integer numbers (not containing duplicates). The most important
list operation is a transformation $\mathcal{T}$ which takes two
lists as input and uses the elements of the former as indexes to
select some elements from the latter. In other words, if $a$ and $b$
are lists, then $\mathcal{T}_a(b)$ is a list whose $i$-th element is
$b_{a_i}$. \par \noindent
\picinpar{
  \quad\quad The following example will show how $\mathcal{T}$ is
  used for matching operators with registers. The time slice in
  the insert on the right is specified by the single index list
  $\ell = (0,2,3)$, and represents the circuit $H \otimes \mathbb{I}
  \otimes H \otimes H$. When it is executed on a quantum register
  with associated address list $r = (r_0, r_1, \dots)$, the language
  run time environment must pair the elements of $\ell$ and $r$,
  forming a new list $\bar{r} = \mathcal{T}_{\ell}(r) = (r_{\ell_0},
  r_{\ell_1}, \dots)$ which in the current example gives $\bar{r}
  = (r_0, r_2, r_3)$.}
{\begin{qcircuit}{15}{1}{4}{10}
    \qlabel{3}{0} \qlabel{2}{1} \qlabel{1}{2} \qlabel{0}{3}
    \qlabel{4}{} 
    \qhadamard{0} \qsame \qhadamard{1} \qsame \qhadamard{3}
  \end{qcircuit}}
The $\mathcal{T}_{\ell}$ transformation corresponds to the ``index
to address translation'' stage for the time slice of Hadamard gates
in figure \ref{fig:translation}. The $\bar{r}$ list is not immediately
sent to the quantum device, for reasons which will be apparent later,
but undergoes a further mapping%
\footnote{This mapping is indeed a permutation since each address must
  correspond to a physical location, and two distinct addresses must
  map to two distinct locations.}
which implements the ``address permutation'' $\mathcal{P}$ in figure
\ref{fig:translation}: $\bar{r} \rightarrow \mathcal{P}(\bar{r})
= (\mathcal{P}_{\bar{r}_0}, \mathcal{P}_{\bar{r}_1}, \dots)$. This
translation, though different in nature from the previous one, can
use the very same algorithm if a list $p = (\mathcal{P}_0,
\mathcal{P}_1, \dots)$ is provided; in this case $\mathcal{P}(\bar{r})$
is equal to $\mathcal{T}_{\bar{r}}(p)$. Summarising, if a time slice
represents a real quantum operation, the addresses which are sent to
the quantum device are $\mathcal{T}_{\mathcal{T}_{\ell}(r)}(p)$ for
each list $\ell$ in the slice. \par
The behaviour of the language run time environment is however different
if the time slice represents a classical permutation, like a ``qubit
line swap'', which is the action of exchanging the quantum state of
two qubits. This action can be implemented by three CNOT gates as
shown in the following circuit decomposition:
\begin{center}
  \begin{qcircuit}{15}{1}{2}{0}
    \qswap{0}{1}
  \end{qcircuit}
  \quad \quad is equivalent to \quad \quad
  \begin{qcircuit}{15}{3}{2}{0}
    \qnot{0}{1} \qnot{1}{0} \qnot{0}{1}
  \end{qcircuit}
\end{center}
This circuit shows that the exchange is a legal quantum operation and
that it can be implemented by sending the appropriate control commands
to the quantum device. It is however also obvious that the additional
mapping $\mathcal{P}$ between the qubit addresses in the quantum
registers and the qubit locations in the quantum device can be used
in order to achieve the same result; in this case it is sufficient
to modify the list $p$ appropriately. \par
This approach is preferable for two reasons. First, it concerns only
the classical machine, hence leading to a smaller number of quantum
operations to be actually performed. Second, it modifies the addresses
which are sent to the quantum device transparently to the registers:
this means that if two registers overlap and one of them undergoes
a number of qubit line swaps, subsequent mappings of the addresses
of the other one are influenced too. Therefore, the programmer can
still think as if the qubit line swap was a real quantum operation.
The additional mapping $\mathcal{P}$ is sufficient to implement the
\texttt{QSwap} quantum operator, described at
pag.\pageref{sec:qops:reorder}. \par\noindent
\picinpar
{\quad\quad In the insert on the right a time slice for a qubit line
  swap operation can be seen; this slice needs two index lists, the
  corresponding elements of which are the line indexes to be exchanged.
  The two lists in the example are $\ell^{\;(0)} = (0, 1)$ and
  $\ell^{\;(1)} = (3, 2)$. The first stage of address translation is
  the same as before, the two lists are combined with the register to
  form $\bar{r}^{\;(0)} = \mathcal{T}_{\ell^{(\!0\!)}}(r) = (r_0, r_1)$
  and $\bar{r}^{\;(1)} = \mathcal{T}_{\ell^{(\!1\!)}}(r) = (r_3, r_2)$.}
{\begin{qcircuit}{15}{1}{4}{10}
    \qlabel{3}{0} \qlabel{2}{1} \qlabel{1}{2} \qlabel{0}{3} 
    \qsame[0.3] \qswap{0}{3} \qsame[0.6] \qswap{1}{2}
  \end{qcircuit}}
The swaps can then be easily implemented by transposing the
$\bar{r}^{\;(0)}_i$-th and the $\bar{r}^{\;(1)}_i$-th elements
of the list $p$ for each valid $i$; in the example, this means
transposing $p_{r_0}$ with $p_{r_3}$ and $p_{r_1}$ with $p_{r_2}$.
The transposition preserves the property of $\mathcal{P}$ of being
a permutation. Summarising, if a time slice represents a qubit line
swap no commands are sent to the quantum device; instead, for each
address pair $(a_i, b_i)$, where $a_i$ is the $i$-th element of
$\mathcal{T}_{\ell^{(\!0\!)}}(r)$ and $b_i$ is the $i$-th element
of $\mathcal{T}_{\ell^{(\!1\!)}}(r)$ , the elements $p_{a_i}$ and
$p_{b_i}$ are transposed in the list $p$.

\clearpage

\newcommand{\code}[1]{\tt #1}
\newcommand{\myline}[3]{#1 & \code{#2} & #3 \\ }
\newcommand{\myhead}[1]{{\rule[-4pt]{0pt}{16pt}\normalsize\bf\em #1}}
\newcommand{\mypage}[1]{p.\pageref{#1}}
\newcommand{\mysec}[1]{sec.\ref{#1}}
\newcommand{\mytable}[4]{
  \begin{table}
    \caption{#1 (#2)} \label{#3} \footnotesize
    \begin{tabular}{|p{.25\linewidth}|p{.55\linewidth}|p{.08\linewidth}|}
      \hline  & \myhead{Prototype} & \myhead{Ref.} \\
      \hline #4 \hline
    \end{tabular}
  \end{table}}

\mytable{Quantum registers, the {\tt Qreg} objects}
{see section \ref{sec:regs}} {tab:regs}
{
  \myline{The register class}{class Qreg;}{\mysec{sec:regs}}
  \myline{Type for a qubit address}{
    Qreg::address}{\mysec{sec:implementation}}
  \myline{Type for a register size}{
    Qreg::size\_type}{\mysec{sec:qram_model}}
  \myline{Type for a bit set}{Qbitset \textrm{~or~} unsigned integers}{
    \mypage{sec:lowlevel}}
  \hline
  \myline{Register constructors}{
    Qreg::Qreg(size\_type s = 1, value v = 0);}{
    \mypage{sec:regs:alloc}, \pageref{sec:lowlevel:assign}}
  \myline{}{Qreg::Qreg(const Qbitset \&the\_bits);}{
    \mypage{sec:regs:alloc}, \pageref{sec:lowlevel:assign}}
  \myline{Register assignment}{
    void Qreg::operator=(value v) const;}{\mypage{sec:lowlevel:assign}}
  \myline{}{void Qreg::operator=%
    (const Qbitset \&the\_bits) const;}{\mypage{sec:lowlevel:assign}}
  \myline{Measurement (blocking)}{Qbitset %
    Qreg::measure(void) const;}{\mypage{sec:lowlevel:measure}}
  \hline
  \myline{Register copy constructor}{Qreg::Qreg(const Qreg \&a\_register);}{}
  \myline{Register destructor}{Qreg::\~{}Qreg();}{\mypage{sec:regs:dealloc}}
  \hline
  \myline{Qubit addressing}
  {Qreg Qreg::operator[](address a) const;}{\mypage{sec:regs:addconc}}
  \myline{}
  {Qreg Qreg::operator()(address a, size\_type s) const;}
  {\mypage{sec:regs:addconc}}
  \myline{Register concatenation}{Qreg operator\&%
    (const Qreg \&r\_1, const Qreg \&r\_2);}{\mypage{sec:regs:addconc}}
  \myline{}{Qreg \&Qreg::operator\&=%
    (const Qreg \&second\_register);}{\mypage{sec:regs:addconc}}
  \myline{Register resizing}{
    Qreg \&Qreg::operator+=(size\_type the\_size);}{\mypage{sec:regs:resize}}
  \myline{}{
    Qreg \&Qreg::operator-=(size\_type the\_size);}{\mypage{sec:regs:resize}}
  \myline{Register size}{
    Qreg::size\_type Qreg::size(void) const;}{\mypage{sec:regs:alloc}}
}

\mytable{Quantum operators, the {\tt Qop} objects}
{see section \ref{sec:qops}} {tab:qops}
{
  \myline{The operator class}{class Qop;}{\mysec{sec:qops}}
  \hline
  \myline{Default constructor}{Qop::Qop();}{\mypage{sec:qops:ctors}}
  \myline{Copy constructor}{Qop::Qop(const Qop \&op);}{}
  \myline{Controlled operators}{
    Qop::Qop(const Qop \&op, size\_type ctrl);}{\mypage{sec:qops:ctrl}}
  \myline{Oracle operators}{Qop::Qop(int(*f)(int), %
    size\_type in, size\_type out);}{\mysec{sec:pseudoclassical}}
  \myline{Phase oracle operators}{
    Qop::Qop(bool(*f)(int), size\_type in);}{\mysec{sec:pseudoclassical}}
  \hline
  \myline{Operator composition}{Qop operator\&%
    (const Qop \&op\_1, const Qop \&op\_2);}{\mypage{sec:qops:comp}}
  \myline{}{Qop \&Qop::operator\&=(const Qop \&op);}{\mypage{sec:qops:comp}}
  \myline{}{Qop \&Qop::operator<<(Qop \&op);}{\mypage{sec:qops:comp}}
  \myline{Operator conjugation, mutable}{
    Qop \&Qop::adjoin(void);}{\mypage{sec:qops:conj}}
  \myline{Operator conjugation, const}{
    Qop Qop::operator!(void) const;}{\mypage{sec:qops:conj}}
  \myline{Operator split, mutable}{Qop \&Qop::split%
    (size\_type head, size\_type jump);}{\mypage{sec:qops:perm}}
  \myline{Operator invert, mutable}{Qop \&Qop::invert%
    (size\_type head, size\_type size);}{\mypage{sec:qops:perm}}
  \myline{Operator split/invert, const}{Qop Qop::operator()%
    (size\_type,size\_type,op\_type) const;}{\mypage{sec:qops:perm}}
  \myline{Operator offset, mutable}{Qop \&Qop::offset%
    (size\_type jump);}{\mypage{sec:qops:perm}}
  \myline{Operator offset, const}{
    Qop Qop::operator>>(size\_type jump) const;}{\mypage{sec:qops:perm}}
  \hline
  \myline{Operator application}{void Qop::operator()%
    (const Qreg \&a\_register) const;}{\mypage{sec:qops:exec}}
}

\mytable{Computational primitives}
{see section \ref{sec:qops} and \ref{sec:lowlevel}} {tab:cprim}
{
  \myline{Hadamard mixing}{class QHadamard;}{\mypage{sec:lowlevel:prim},$H$}
  \myline{Phase shift($Z$-rotation)}{class QPhase;}{%
    \mypage{sec:lowlevel:prim},$R_k$}
  \myline{Conditional phase shift}{class QCondPhase;}{%
    \mypage{sec:lowlevel:prim},$C_{R_k}$}
  \myline{Controlled NOT}{class QCnot;}{\mypage{ctor:cnot}}
  \myline{Toffoli gate}{class QToffoli;}{\mypage{ctor:toffoli}}
  \myline{Swap gate (classical)}{class QSwap;}{app.\ref{app:translation}}
  \myline{Discrete Fourier transform}{class QFourier;}{\cite{QFT_orig}}
}

\newcommand{\mybibitem}[5][{}]{\bibitem{#2}{#3}, ``{#4}'',{#1} {#5}}
\newcommand{\journal}[4]{{\em #1}, {\bf #2} (#3) pp.#4}
\newcommand{\proceeding}[3]{{\em Proc. of the #1}, (#2) pp.#3}
\newcommand{\book}[3]{#1, #2, \bf #3}
\newcommand{\preprint}[1]{{\tt #1}}
\newcommand{\web}[1]{{\tt http://#1}}

\end{document}